\documentclass[english, onecolumn,12pt]{IEEEtran}
\usepackage[T1]{fontenc}
\usepackage[utf8]{inputenc}
\usepackage{babel}
\usepackage{mathrsfs}
\usepackage{amsmath}
\usepackage{amsthm}
\usepackage{amssymb}
\usepackage{stmaryrd}
\usepackage{graphicx}
\usepackage{geometry}
\usepackage[pdfusetitle,
 bookmarks=true,bookmarksnumbered=false,bookmarksopen=false,
 breaklinks=false,pdfborder={0 0 0},pdfborderstyle={},backref=false,colorlinks=false]
 {hyperref}

\usepackage{hyperref}
\hypersetup{
    hidelinks=true  
}

\makeatletter
\@ifundefined{date}{}{\date{}}
\usepackage{babel}

\providecommand{\definitionname}{Definition}
\providecommand{\lemmaname}{Lemma}
\providecommand{\notationname}{Notation}
\providecommand{\propositionname}{Proposition}
\providecommand{\theoremname}{Theorem}

\providecommand{\remarkname}{Remark}

\theoremstyle{plain}
\newtheorem{thm}{\protect\theoremname}\theoremstyle{definition}
\newtheorem{notation}[thm]{\protect\notationname}\newtheorem{defn}[thm]{\protect\definitionname}\theoremstyle{plain}
\newtheorem{prop}[thm]{\protect\propositionname}\newtheorem{lem}[thm]{\protect\lemmaname}\theoremstyle{remark}
\newtheorem{rem}[thm]{\protect\remarkname}

\makeatother

\providecommand{\definitionname}{Definition}
\providecommand{\lemmaname}{Lemma}
\providecommand{\propositionname}{Proposition}
\providecommand{\remarkname}{Remark}
\providecommand{\theoremname}{Theorem}

\usepackage{amstext}
\begin{document}
\title{Comparative Study of Q-Learning for State-Feedback LQG Control with
an Unknown Model}
\author{ Mingxiang Liu$^{1}$, Damián Marelli$^{2,\dagger}$, Minyue Fu$^{3}$,
\emph{Life Fellow, IEEE},  and Qianqian Cai$^{4}$ \thanks{$^{1}$Mingxiang Liu is with the School of Automation and Intelligent
Manufacturing, Southern University of Science and Technology, Shenzhen,
China. Email: \texttt{liumx@sustech.edu.cn}}\thanks{$^{2}$Damián Marelli is with the School of Automation, Guangdong
University of Technology, Guangzhou, China, and with the French Argentine
International Center for Information and Systems Sciences, National
Scientific and Technical Research Council, Argentina. Email: \texttt{Damian.Marelli@newcastle.edu.au}}\thanks{$^{3}$Minyue Fu is with the School of Automation and Intelligent
Manufacturing, Southern University of Science and Technology, Shenzhen,
China. Email: \texttt{fumy@sustech.edu.cn}}
\thanks{$^{4}$Qianqian Cai is with the School of Automation, Guangdong University
of Technology, Guangzhou, China. Email: \texttt{qianqian.cai@outlook.com}}
\thanks{This work was supported by the Argentinean Agency for Scientific and
Technological Promotion (PICT-2021-I-A-00730), the National Foreign
Expert Individual Project (H20240983), the National Natural Science
Foundation of China (62350710214, 62273100, U23A20325, U21A20476),
the Guangdong Basic and Applied Research Foundation (2024A1515010193),
and the Shenzhen Key Laboratory of Control Theory and Intelligent
Systems (ZDSYS20220330161800001).}\thanks{$\dagger$Corresponding author.}}
\maketitle
\begin{abstract}
We study the problem of designing a state feedback linear quadratic
Gaussian (LQG) controller for a system in which the system matrices
as well as the process noise covariance are unknown. We do a rigorous
comparison between two approaches. The first is the classic one in
which a system identification stage is used to estimate the unknown
parameters, which are then used in a state-feedback LQG (SF-LQG) controller
design. The second approach is a recently proposed one using a reinforcement
learning paradigm called Q-learning. We do the comparison in terms
of complexity and accuracy of the resulting controller. We show that
the classic approach asymptotically efficient, giving virtually no
room for improvement in terms of accuracy. We also propose a novel
Q-learning-based method which we show asymptotically achieves the
optimal controller design. We complement our proposed method with
a numerically efficient algorithmic implementation aiming at making
it competitive in terms of computations. Nevertheless, our complexity
analysis shows that the classic approach is still numerically more
efficient than this Q-learning-based alternative. We then conclude
that the classic approach remains being the best choice for addressing
the SF-LQG design in the case of unknown parameters. 
\end{abstract}


\section{Introduction}

Reinforcement Learning (RL) has achieved remarkable success across
a wide array of applications, particularly in tackling sequential
decision-making problems~\cite{mnih2015human,silver2016mastering}.
Within the control community, dynamic systems are often modeled as
Markov decision processes, with RL techniques employed to solve optimal
control problems. RL approaches in control can be classified into
two main categories: direct policy optimization via policy search
and value function approximation through dynamic programming~\cite{recht2018tour,hu2023theoretical,lewis2009adaptive,jiang2020learning}.
The latter can in turn be classified in methods based in either value
iteration or policy iteration~\cite{bradtke1994adaptive,landelius1997reinforcement,jiang2012computational,CTVI}.
These have been extensively studied and applied to deterministic linear
systems.

Despite these advancements, practical engineering applications often
involve the challenge of estimating the optimal control law using
noisy data. Addressing this issue is crucial yet relatively underexplored.
Works within this line can be grouped along continuous-time (CT) and
discrete-time (DT) lines. For CT systems,~\cite{pang2021robust,pang2022reinforcement}
analyze the robustness of policy iteration under additive disturbances
as well as joint multiplicative-and-additive noise. For DT systems,
in~\cite{guo2023lqg}, the LQG problem is addressed using supervised
learning, which requires the availability of \textit{\emph{a dataset
of optimal input--output}} samples. The RL approach is instead used
in~\cite{yaghmaie2022lqg}, where the authors show convergence of
the estimated controller, but without optimality guarantees. Also,~\cite{cui2024risk}
uses RL to design a controller aiming at minimizing a criterion called
linear quadratic zero-sum dynamic game rather than the LQG one. This
design is proposed without convergence guarantees. A further limitation
in several approaches is the reliance on a priori knowledge of noise
statistics: for example, \cite{krauth2019finite} introduces a Q-learning
method to solve the LQG problem, and \cite{lai2023modelfree} proposes
a policy iteration algorithm tailored for systems affected by both
additive and multiplicative noise, but both methods require prior
knowledge of the additive noise covariance. This limits their applicability
in scenarios where such information is unavailable.

In an effort to overcome these limitations,~\cite{kumar2023unconstrained}
presents a value iteration algorithm that does not require knowledge
of the system matrices or the noise covariance. Nevertheless, this
approach lacks a formal convergence proof. Similarly, in~\cite{jiang2024adaptive},
the authors propose an algorithm to address the linear-quadratic control
problem in the presence of noise. In this approach, iterations proceed
with a fixed horizon of collected data. Hence, as the number of iterations
increases, the error stabilizes and ceases to decrease any further.

Furthermore, a more fundamental question remains unaddressed in the
literature. While these novel RL approaches attempt to bypass explicit
system identification, it is not clear whether they offer any tangible
advantage in accuracy or computational complexity over the classic
approach. A rigorous benchmark comparison for the unknown-parameter
SF-LQG problem is currently lacking.

In this paper, we aim to address both of these identified gaps. We
first continue the research line of~\cite{kumar2023unconstrained,jiang2024adaptive}
to develop a provably convergent algorithm, and then use it to conduct
the aforementioned benchmark comparison. The contributions of our
work are as follows:

In this paper, we aim to address both of these identified gaps. We
first continue the research line of~\cite{kumar2023unconstrained,jiang2024adaptive}
to develop a provably convergent algorithm, and then use it to conduct
the aforementioned benchmark comparison. The contributions of our
work are as follows:
\begin{enumerate}
\item We propose a Q-learning algorithm designed for SF-LQG problems without
prior knowledge of the system matrix and the process noise covariance.
In comparison with the algorithms in~\cite{kumar2023unconstrained,jiang2024adaptive},
the advantage of our algorithm is that it is guaranteed to approximate
the optimal controller in the sense that the approximation error approaches
zero as the number of iterations increases. We provide a rigorous
proof of this property. 
\item The classic approach for solving the problem consists in combining
a system identification stage to estimate the unknown system parameters,
together with a SF-LQG stage to design the controller using the estimated
parameters. We do a rigorous performance comparison between the classic
approach and our proposed Q-learning method. The comparison is done
in terms of both, accuracy and complexity. The following two contributions
are the steps towards this comparison. 
\item Aiming at the accuracy comparison, we show that, if the system identification
stage of the classic approach is done using the maximum likelihood
criterion, the resulting controller design asymptotically reaches
the theoretical lower accuracy bound given by the Cramér Rao lower
bound (CRLB). An immediate consequence of this result is that virtually
no alternative approach can compete with the classic one in terms
of accuracy. 
\item Aiming at the complexity comparison, we derive a numerically efficient
implementation of our proposed Q-learning algorithm. Using this implementation
we conduct a detailed complexity analysis of both approaches. We conclude
that the classic approach outperforms the Q-learning approach also
in terms of complexity. This makes us conclude that there may not
be any advantage in using Q-learning techniques for solving the SF-LQG
problem in the case of unknown system parameters. 
\end{enumerate}
The rest of the paper is organized as follows: In Section~\ref{sec:Problem-description}
we introduce the research problem. In Section~\ref{sec:Preliminary-result}
we derive a result which is instrumental for our work, concerning
the discrete algebraic Riccati equation (DARE). In Section~\ref{sec:Classic-solution}
we conduct our study on the classic approach, including its accuracy
and complexity. In Section~\ref{sec:Solution-using-Q-learning} we
do the same for the Q-learning approach. In Section~\ref{sec:Simulations}
we present numerical simulations and give concluding remarks in Section~\ref{sec:Conclusion}.
\begin{notation} We use $\llbracket T\rrbracket=\text{\ensuremath{\left\{ t:1\leq t\leq T\right\} } }.$
For a sequence $\left(x_{t}:t\in\mathbb{N}_{0}\right)$ (where $\mathbb{N}_{0}=\{0\}\cup\mathbb{N}$)
and $0\leq a\leq b\leq\infty$,we use the notation 
\[
X_{a,b}=\left(x_{t}:a\leq t\leq b\right).
\]
For a set of (possibly vector) random variables $X_{a,b}=\left(x_{t}:a\leq t\leq b\right),$
we use $\sigma\left(X_{a,b}\right)$ to denote the $\sigma$-algebra
generated by the random variables $\left(x_{t}:a\leq t\leq b\right)$.
For a random variable $y$ and $\sigma$-algebra $\mathcal{A}$, we
use $y\in\mathcal{A}$ to denote that $\sigma(y)\subseteq\mathcal{A}$,
i.e., $y$ is measurable with respect to $\mathcal{A}$. Also, 
\[
\mathcal{F}\left(X_{a,b}\right)=\left(\sigma\left(X_{a,t}\right):a\leq t\leq b\right)
\]
denotes the natural filtration generated by $X_{a,b}$. For $c\geq a$
and $d\leq b$, we say that $Y_{c,d}=\left(y_{t}:c\leq t\leq d\right)$
is adapted to $X_{a,b}$, and denote it by 
\[
Y_{c,d}\in\mathcal{F}\left(X_{a,b}\right),
\]
if $y_{t}\in\sigma\left(X_{a,t}\right)$, for all $c\leq t\leq d$.
Intuitively, the latter means that $y_{t}$ is a function of $X_{a,t}$.
\end{notation} For a symmetric matrix $X\in\mathbb{R}^{N\times N}$
we define the upper triangular vectorization operation $\overset{\rightsquigarrow}{X}$
by 
\begin{align*}
\overset{\rightsquigarrow}{X}^{\top} & =\left[X^{\top}_{1},\cdots,X^{\top}_{N}\right],\\
X^{\top}_{n} & =\left[\left[X\right]_{1,n},\cdots,\left[X\right]_{n,n}\right].
\end{align*}
For vectors $a,b\in\mathbb{R}^{N}$, we also define the upper triangular
Kronecker product $\tilde{\otimes}$ by 
\begin{align*}
\left(a\tilde{\otimes}b\right)^{\top} & =\left[c^{\top}_{1},\cdots,c^{\top}_{N}\right],\\
c^{\top}_{n} & =\left[a_{1}b_{n}+a_{n}b_{1},\cdots,a_{n-1}b_{n}+a_{n}b_{n-1},a_{n}b_{n}\right].
\end{align*}
Notice that with the above definitions 
\[
a^{\top}Xb=\left(b\tilde{\otimes}a\right)^{\top}\overset{\rightsquigarrow}{X}.
\]

\section{Problem description}

\label{sec:Problem-description}

\paragraph{The SF-LQG problem}

Consider the following state-transition equation 
\begin{equation}
x_{t+1}=Ax_{t}+Bu_{t}+w_{t},\label{eq:system}
\end{equation}
where $x_{0}\in\mathbb{R}^{N}$ is the initial state, $u_{t}\in\mathbb{R}^{M}$
is the input and $w_{t}\sim\mathcal{N}(0,\Sigma)$ is the process
noise, which we assume to be an independent sequence. We also assume
that $A$ is non-singular and $(A,B)$ is stabilizable. The SF-LQG
problem consists in solving, at each $T\in\mathbb{N}_{0}$, the following
optimization problem 
\begin{equation}
U^{\star}_{T,\infty}\left(x_{T}\right)=\underset{U_{T,\infty}\in\mathcal{F}\left(X_{0,\infty}\right)}{\arg\min}J_{T}\left(x_{T},U_{T,\infty}\right),\label{eq:SF-LQG-problem}
\end{equation}
where 
\[
J_{T}\left(x_{T},U_{T,\infty}\right)=\mathcal{E}\left\{ \sum^{\infty}_{t=0}\gamma^{t}\left(\left\Vert x_{t}\right\Vert ^{2}_{Q}+\left\Vert u_{t}\right\Vert ^{2}_{R}\right)\right\} ,
\]

for some $Q,R>0$ (i.e., $Q$ and $R$ are symmetric positive definite
matrices). Also, $\mathcal{E}$ denotes the expectation operation.
Let $U^{\star}_{T,\infty}\left(x_{t}\right)=\left(u^{\star}_{t}\left(x_{T}\right):t\leq s<\infty\right)$.
Then, at time $T$, the SF-LQG controller discards all future computed
inputs $u^{\star}_{t}\left(x_{T}\right):T<t<\infty$, and puts 
\begin{equation}
u_{T}=u^{\star}_{T}\left(x_{T}\right).\label{eq:MPC}
\end{equation}

\paragraph{The SF-LQG solution}

The classic solution to the SF-LQG problem is given by~\cite{1970Introduction}
\begin{equation}
u^{\star}_{T}\left(x_{T}\right)=-Lx_{T},\label{eq:control-policy}
\end{equation}
where 
\begin{equation}
L=\gamma\left(\gamma B^{\top}PB+R\right)^{-1}B^{\top}PA,\label{eq:Ldef}
\end{equation}
and $P$ is the unique positive solution (see Proposition~\ref{pro:DARE}
for a justification of this claim) of the following DARE 
\begin{equation}
P=\gamma A^{\top}PA-\gamma^{2}A^{\top}PB\left(\gamma B^{\top}PB+R\right)^{-1}B^{\top}PA+Q.\label{eq:Pdef}
\end{equation}
Moreover 
\begin{equation}
J_{T}\left(x_{T},U^{\star}_{T,\infty}\left(x_{T}\right)\right)=x^{\top}_{T}Px_{T}+\frac{\gamma}{1-\gamma}\mathrm{Tr}\left\{ \Sigma P\right\} .\label{eq:SF-LQG-Jopt}
\end{equation}

\paragraph{Research problem}

Solving the SF-LQG problem requires knowledge of the system parameters
$A$, $B$ and $\Sigma$. We assume that these parameters are not
known, and instead we know, at time $T$, all past inputs $U_{0,T-t}$
and states $X_{0,T}$. The classic approach for exploiting this knowledge
consists in estimating the system parameters using system identification
techniques, and then use these estimates to design a controller using~(\ref{eq:control-policy})-(\ref{eq:Pdef}).
An alternative to this approach, based on the Q-learning algorithm,
has recently been proposed~\cite{kumar2023unconstrained}. It essentially
consists in directly estimating, at each time $t$, the control strategy~(\ref{eq:control-policy})
using the available data $U_{0,T-t}$ and $X_{0,T}$.

Our goal is to compare these two approaches in terms of accuracy and
computational complexity.

In view of~(\ref{eq:SF-LQG-Jopt}), the performance of the SF-LQG
controller depends on matrix $P$. Hence, as accuracy index we use
the following estimation error covariance 
\begin{equation}
\mathcal{C}_{T}=\mathcal{E}\left\{ \left(\overset{\rightsquigarrow}{P}-\overset{\rightsquigarrow}{\hat{P}}_{T}\right)\left(\overset{\rightsquigarrow}{P}-\overset{\rightsquigarrow}{\hat{P}}_{T}\right)^{\top}\right\} ,\label{eq:error-covariance}
\end{equation}
(recall that $\overset{\rightsquigarrow}{\cdot}$ denotes the vectorization
operation applied to a symmetric matrix) where $\hat{P}_{T}$ denotes
the estimate of $P$ obtained by the estimation stage of the studied
method. We say that $\hat{P}_{T}$ is strongly consistent if it converges,
with probability one, to the true value $P$. We also say that it
is asymptotically efficient if $T\mathcal{C}_{T}$ converges to its
lower theoretical bound, i.e., the CRLB.

\section{Preliminary result: solution to the discrete algebraic Riccati equation}

\label{sec:Preliminary-result}

In this section we introduce a result which is instrumental to the
two methods studied in the paper. This consists in providing conditions
for the convergence, as well as its rate, of the recursions resulting
from iterating a DARE.

We start by introducing a required concept, namely, the Thomson distance
between positive-definite matrices. \begin{notation} For a given
$N\in\mathbb{N}$, let $\mathbb{P}_{N}\left(\mathbb{R}\right)$ denote
the set of positive-definite matrices. \end{notation} The following
definition introduces a metric for the space $\mathbb{P}_{N}\left(\mathbb{R}\right)$. 
\begin{defn}
For $P,Q\in\mathbb{P}_{N}\left(\mathbb{R}\right)$, the Thomson distance
between $P$ and $Q$ is given by 
\[
\delta\left(P,Q\right)=\left\Vert \log Q^{-1/2}PQ^{-1/2}\right\Vert _{\mathrm{F}},
\]
where $\|\cdot\|_{\mathrm{F}}$ denotes the Frobenius norm. 
\end{defn}
It is shown in~\cite[Chapter 6]{bhatia2009positive} that $\delta$
is indeed a metric. As shown in~\cite{marelli2021distributed}, $\delta$
also enjoys the following properties. 
\begin{prop}
\label{prop:thomson}For $P,Q,R\in\mathbb{P}_{N}\left(\mathbb{R}\right)$: 
\begin{enumerate}
\item $\delta\left(P^{-1},Q^{-1}\right)=\delta\left(P,Q\right)$; 
\item for any $W\geq0$ and $B$ such that $W+BPB^{\top},W+BQB^{\top}\in\mathbb{P}_{N}\left(\mathbb{R}\right),$
\[
\delta\left(W+BPB^{\top},W+BQB^{\top}\right)\leq\frac{\alpha}{\alpha+\beta}\delta\left(P,Q\right),
\]
where $\alpha=\max\left\{ \left\Vert BPB^{\top}\right\Vert ,\left\Vert BQB^{\top}\right\Vert \right\} $
and $\beta=\lambda_{\min}(W)$ (i.e., the smallest eigenvalue of $W$).
Equality holds if $W=0$ and $B$ is invertible. 
\item $\left\Vert P-Q\right\Vert \leq\left(e^{\delta\left(P,Q\right)}-1\right)\min\left\{ \left\Vert P\right\Vert ,\left\Vert Q\right\Vert \right\} $. 
\end{enumerate}
\end{prop}
The following proposition states the main result of the section. 
\begin{prop}
\label{pro:DARE} If $Q,R>0$ and $(A,B)$ is stabilizable, then the
equation 
\begin{equation}
P=A^{\top}PA+Q-A^{\top}PB\left[R+B^{\top}PB\right]^{-1}B^{\top}PA,\label{eq:DARE}
\end{equation}
has a unique positive definite solution $P>0$. Also, for any $P_{0}>0$,
the recursions 
\begin{equation}
P_{k+1}=A^{\top}P_{k}A+Q-A^{\top}P_{k}B\left[R+B^{\top}P_{k}B\right]^{-1}B^{\top}P_{k}A,\label{eq:DARErec}
\end{equation}
converge to $P$, i.e., $\lim_{k\rightarrow\infty}P_{k}=P.$ Moreover,
for all $k\in\mathbb{N}$, 
\[
\left\Vert P-P_{k}\right\Vert \leq\left(e^{\rho^{k}\delta\left(P,P_{0}\right)}-1\right)\left\Vert P\right\Vert ,
\]
where 
\[
\rho=\frac{1}{1+\left(\left\Vert Q^{-1}\right\Vert \left\Vert A\right\Vert ^{2}\left\Vert P\right\Vert \right)^{-1}e^{-\delta\left(P,P_{0}\right)}}.
\]
\end{prop}
\begin{IEEEproof}
We split the proof in steps:

1) For $i=1,2$, let $P^{(i)}_{0}>0$ and 
\begin{equation*}
P^{(i)}_{k+1}=A^{\top}P^{(i)}_{k}A+Q-A^{\top}P^{(i)}_{k}B\left[R+B^{\top}P^{(i)}_{k}B\right]^{-1}B^{\top}P^{(i)}_{k}A.
\end{equation*}
Let 
\begin{equation}
\bar{P}^{(i)}_{k}=P^{(i)}_{k}-P^{(i)}_{k}B\left[R+B^{\top}P^{(i)}_{k}B\right]^{-1}B^{\top}P^{(i)}_{k},\label{eq:update}
\end{equation}
so that 
\[
P^{(i)}_{k+1}=A^{\top}\bar{P}^{(i)}_{k}A+Q.
\]
Let $\Gamma^{(i)}_{k}=\left(P^{(i)}_{k}\right)^{-1}$ and $\bar{\Gamma}^{(i)}_{k}=\left(\bar{P}^{(i)}_{k}\right)^{-1}$.
Using the matrix inversion lemma, we can write~(\ref{eq:update})
as 
\[
\bar{\Gamma}^{(i)}_{k}=\Gamma^{(i)}_{k}+B^{\top}R^{-1}B.
\]
Then, using Proposition~\ref{prop:thomson}-1 and~2 
\begin{align}
\delta\left(P^{(1)}_{k+1},P^{(2)}_{k+1}\right) & =\delta\left(A\bar{P}^{(1)}_{k}A^{\top}+Q,A\bar{P}^{(2)}_{k}A^{\top}+Q\right)\nonumber \\
 & \leq\frac{\alpha_{1}}{\alpha_{1}+\beta_{1}}\delta\left(\bar{P}^{(1)}_{k},\bar{P}^{(2)}_{k}\right),\label{eq:contraction1}
\end{align}
with 
\begin{align*}
\alpha_{1} & =\max\left\{ \left\Vert A\bar{P}^{(1)}_{k}A^{\top}\right\Vert ,\left\Vert A\bar{P}^{(2)}_{k}A^{\top}\right\Vert \right\} ,\\
\beta_{1} & =\lambda_{\min}\left(Q\right).
\end{align*}
Also 
\begin{equation}
	\begin{aligned}
\delta\left(\bar{P}^{(1)}_{k},\bar{P}^{(2)}_{k}\right)=\delta\left(\bar{\Gamma}^{(1)}_{k},\bar{\Gamma}^{(2)}_{k}\right)
&=\delta\left(\Gamma^{(1)}_{k}+B^{\top}R^{-1}B,\Gamma^{(2)}_{k}+B^{\top}R^{-1}B\right)\\
&\leq\frac{\alpha_{2}}{\alpha_{2}+\beta_{2}}\delta\left(\Gamma^{(1)}_{k},\Gamma^{(2)}_{k}\right),\label{eq:contraction2}
	\end{aligned}
\end{equation}
with 
\begin{align*}
\alpha_{2} & =\max\left\{ \left\Vert \Gamma^{(1)}_{k}\right\Vert ,\left\Vert \Gamma^{(2)}_{k}\right\Vert \right\} ,\\
\beta_{2} & =\lambda_{\min}\left(B^{\top}R^{-1}B\right).
\end{align*}
From~(\ref{eq:contraction1}) and~(\ref{eq:contraction2}) 
\begin{equation}
\delta\left(P^{(1)}_{k+1},P^{(2)}_{k+1}\right)\leq\rho\delta\left(P^{(1)}_{k},P^{(2)}_{k}\right),\label{eq:contraction}
\end{equation}
with 
\[
\rho=\frac{\alpha_{1}}{\alpha_{1}+\beta_{1}}\frac{\alpha_{2}}{\alpha_{2}+\beta_{2}}.
\]
Since $Q>0$, then $\rho<1$.

2) Since $Q>0$, the pair $\left(Q,A\right)$ is observable. It then
follows from~\cite[Theorem 13.1.3]{lancaster1995algebraic} that~(\ref{eq:DARE})
has a positive solution $P$. To show uniqueness, notice that if $P^{(1)}$
and $P^{(2)}$ are two solutions of~(\ref{eq:DARE}), from~(\ref{eq:contraction})
we obtain 
\[
\delta\left(P^{(1)},P^{(2)}\right)\leq\rho\delta\left(P^{(1)},P^{(2)}\right).
\]
Since $\rho<1$, the above can only hold if $P^{(1)}=P^{(2)}$.

3) For the convergence of~(\ref{eq:DARErec}), from~(\ref{eq:contraction})
we obtain 
\[
\delta\left(P,P_{k+1}\right)\leq\rho_{k}\delta\left(P,P_{k}\right),
\]
with 
\[
\rho_{k}\leq\frac{\alpha_{1,k}}{\alpha_{1,k}+\beta_{1}}.
\]
Now 
\begin{align}
\alpha_{1,k} & =\max\left\{ \left\Vert A\bar{P}A^{\top}\right\Vert ,\left\Vert A\bar{P}_{k}A^{\top}\right\Vert \right\} \nonumber \\
 & \leq\left\Vert A\right\Vert ^{2}\max\left\{ \left\Vert \bar{P}\right\Vert ,\left\Vert \bar{P}_{k}\right\Vert \right\} ,\label{eq:alpha2k}
\end{align}
with 
\[
\bar{P}=P-PB\left[R+B^{\top}PB\right]^{-1}B^{\top}P,
\]
Also, since $\rho_{k}<1$, for all $k\in\mathbb{N}$, we obtain from
Proposition~\ref{prop:thomson}-3 
\begin{align*}
\left\Vert \bar{P}-\bar{P}_{k}\right\Vert  & \leq\left(e^{\delta\left(\bar{P},\bar{P}_{k}\right)}-1\right)\left\Vert \bar{P}\right\Vert \\
 & \leq\left(e^{\delta\left(\bar{P},\bar{P}_{0}\right)}-1\right)\left\Vert \bar{P}\right\Vert \\
 & \leq\left(e^{\delta\left(P,P_{0}\right)}-1\right)\left\Vert \bar{P}\right\Vert .
\end{align*}
Then 
\begin{align*}
\left\Vert \bar{P}_{k}\right\Vert  & \leq\left\Vert \bar{P}\right\Vert +\left\Vert \bar{P}-\bar{P}_{k}\right\Vert \\
 & \leq e^{\delta\left(P,P_{0}\right)}\left\Vert \bar{P}\right\Vert .
\end{align*}
Putting the above into~(\ref{eq:alpha2k}) 
\begin{align*}
\alpha_{1,k} & \leq\left\Vert A\right\Vert ^{2}\left\Vert \bar{P}\right\Vert \max\left\{ 1,e^{\delta\left(P,P_{0}\right)}\right\} \\
 & \leq\left\Vert A\right\Vert ^{2}\left\Vert P\right\Vert e^{\delta\left(P,P_{0}\right)}.
\end{align*}
Then 
\[
\rho_{k}\leq\frac{1}{1+\left(\left\Vert Q^{-1}\right\Vert \left\Vert A\right\Vert ^{2}\left\Vert P\right\Vert \right)^{-1}e^{-\delta\left(P,P_{0}\right)}}\triangleq\rho.
\]
The above implies that 
\[
\delta\left(P,P_{k}\right)\leq\rho^{k}\delta\left(P,P_{0}\right).
\]
In view of Proposition~\ref{prop:thomson}-3 
\begin{align*}
\left\Vert P-P_{k}\right\Vert  & \leq\left(e^{\delta\left(P,P_{k}\right)}-1\right)\left\Vert P\right\Vert \\
 & \leq\left(e^{\rho^{k}\delta\left(P,P_{0}\right)}-1\right)\left\Vert P\right\Vert .
\end{align*}
\end{IEEEproof}

\section{Classic solution using system identification}

\label{sec:Classic-solution}

In this section we describe the classic approach which consists in
combining a system identification stage with the SF-LQG design described
in Section~\ref{sec:Problem-description}. In Section~\ref{subsec:Identification-using-maximum}
we describe the system identification stage. In Section~\ref{subsec:Asymptotic-properties-sysid}
we show that the resulting estimate is both strongly consistent and
asymptotically efficient. Finally, in Section~\ref{subsec:Complexity-analysis-sysid}
we study the complexity of the resulting algorithm.

\subsection{Identification using maximum likelihood}

\label{subsec:Identification-using-maximum}

To estimate the unknown parameters $A$, $B$ and $\Sigma$ we use
the maximum likelihood criterion. This gives 
\begin{equation}
\left\{ \hat{A}_{T},\hat{B}_{T},\hat{\Sigma}_{T}\right\} \in\underset{\left\{ A,B,\Sigma\right\} }{\arg\max}~p\left(X_{0,T};A,B,\Sigma\right).\label{eq:ML-problem}
\end{equation}
Notice that the right-hand side of~(\ref{eq:ML-problem}) is a set,
hence the left-hand side is any member of this set. A solution to
the above problem is given in the next result: \begin{notation} Let
\begin{align*}
\xi^{\top}_{t} & =\left[x^{\top}_{t},u^{\top}_{t}\right],\\
\zeta_{t}(A,B) & =x_{t+1}-\left[A,B\right]\xi_{t},
\end{align*}
and 
\begin{align*}
U_{T} & =\sum^{T}_{t=1}\xi_{t}\xi^{\top}_{t},\\
V_{T} & =\sum^{T}_{t=1}\xi_{t}x^{\top}_{t+1},\\
W_{T}(A,B) & =\frac{1}{T}\sum^{T}_{t=1}\zeta_{t}(A,B)\zeta^{\top}_{t}(A,B).
\end{align*}
\end{notation} 
\begin{thm}
\label{thm:ML} A solution of~(\ref{eq:ML-problem}) is given by
\begin{align}
\left[\hat{A}_{T},\hat{B}_{T}\right]^{\top} & =U^{\dagger}_{T}V_{T},\label{eq:ML-AB}\\
\hat{\Sigma}_{T} & =W_{T}\left(\hat{A}_{T},\hat{B}_{T}\right),\label{eq:ML-Sigma}
\end{align}
where $^{\dagger}$ denotes the Moore-Penrose pseudoinverse. 
\end{thm}
\begin{IEEEproof}
We split the proof in steps:

1) We have 
\[
p\left(X_{1,T+1};A,B,\Sigma\right)=\prod^{T}_{t=1}p\left(x_{t+1}|x_{t};A,B,\Sigma\right).
\]
Let 
\[
l(A,B,\Sigma)=-\frac{2}{T}\log p\left(X_{1,T+1};A,B,\Sigma\right).
\]
Then 
\begin{align*}
 & l(A,B,\Sigma)\\
= & -\frac{2}{T}\sum^{T}_{t=1}\log p\left(x_{t+1}|x_{t};A,B,\Sigma\right)\\
= & \frac{1}{T}\log(2\pi)^{N}+\frac{1}{T}\log\det\Sigma+\\
 & +\frac{1}{T}\sum^{T}_{t=1}\left(x_{t+1}-Ax_{t}-Bu_{t}\right)^{\top}\Sigma^{-1}\times\\
 & \times\left(x_{t+1}-Ax_{t}-Bu_{t}\right)\\
= & \frac{1}{T}\log(2\pi)^{N}+\frac{1}{T}\log\det\Sigma\\
 & +\mathrm{Tr}\left\{ \Sigma^{-1}\frac{1}{T}\sum^{T}_{t=1}\zeta_{t}(A,B)\zeta^{\top}_{t}(A,B)\right\} \\
= & \frac{1}{T}\log(2\pi)^{N}+\frac{1}{T}\log\det\Sigma+\mathrm{Tr}\left\{ \Sigma^{-1}W_{T}(A,B)\right\} .
\end{align*}

2) We use $\mathfrak{D}$ to denote Fréchet derivatives. Its definition
and properties are summarized in Appendix~\ref{sec:Frechet derivatives}.
From Proposition~\ref{prop:FD-properties}-\ref{enu:Chain-rule})
we obtain 
\begin{equation}
\mathfrak{D}_{[A,B]}l(A,B,\Sigma)(\mathfrak{M})
=\mathfrak{D}_{W_{T}(A,B)}F\left(W_{T}(A,B)\right)\left(\mathfrak{D}_{[A,B]}\left\{ W_{T}(A,B)\right\} (\mathfrak{M})\right),\label{eq:aux-8}
\end{equation}
where 
\begin{align*}
F\left(W_{T}(A,B)\right)
=\frac{1}{T}\log(2\pi)^{N}+\frac{1}{T}\log\det\Sigma+\mathrm{Tr}\left\{ \Sigma^{-1}W_{T}(A,B)\right\} .
\end{align*}
Now, using Lemma~\ref{lem:FD-useful}-\ref{enu:trace}) 
\begin{equation}
\mathfrak{D}_{W_{T}(A,B)}F\left(W_{T}(A,B)\right)(\mathfrak{M})=\mathrm{Tr}\left\{ \Sigma^{-1}\mathfrak{M}\right\} .\label{eq:aux-7}
\end{equation}
Also, from Proposition~\ref{prop:FD-properties}-\ref{enu:Linearity}),
\begin{equation}
\mathfrak{D}_{[A,B]}\left\{ M_{T}\right\} (\mathfrak{M})=\frac{1}{T}\sum^{T}_{t=1}\mathfrak{D}_{[A,B]}\left\{ \zeta_{t}\zeta^{\top}_{t}\right\} (\mathfrak{M}).\label{eq:aux-6}
\end{equation}
Using Proposition~\ref{prop:FD-properties}-\ref{enu:Chain-rule}),
\begin{equation}
\mathfrak{D}_{[A,B]}\left\{ \zeta_{t}\zeta^{\top}_{t}\right\} (\mathfrak{M})=\mathfrak{D}_{\zeta_{t}}\left\{ \zeta_{t}\zeta^{\top}_{t}\right\} \left(\mathfrak{D}_{[A,B]}\left\{ \zeta_{t}\right\} (\mathfrak{M})\right)\label{eq:aux-5}
\end{equation}
From Lemma~\ref{lem:FD-useful}-\ref{enu:outer-prod}) 
\[
\mathfrak{D}_{\zeta_{t}}\left\{ \zeta_{t}\zeta^{\top}_{t}\right\} (\mathfrak{m})=\mathfrak{m}\zeta^{\top}_{t}+\zeta_{t}\mathfrak{m}^{\top},
\]
and from Lemma~\ref{lem:FD-useful}-\ref{enu:affine}) 
\[
\mathfrak{D}_{[A,B]}\left\{ \zeta_{t}\right\} (\mathfrak{M})=-\mathfrak{M}\xi_{t}.
\]
So~(\ref{eq:aux-5}) becomes 
\begin{align*}
\mathfrak{D}_{[A,B]}\left\{ \zeta_{t}\zeta^{\top}_{t}\right\} (\mathfrak{M}) & =\left(-\mathfrak{M}\xi_{t}\right)\zeta^{\top}_{t}+\zeta_{t}\left(-\mathfrak{M}\xi_{t}\right)^{\top}\\
 & =-\mathfrak{M}\xi_{t}\zeta^{\top}_{t}-\zeta_{t}\xi^{\top}_{t}\mathfrak{M}^{\top}.
\end{align*}
Putting the above into~(\ref{eq:aux-6}), 
\begin{align*}
\mathfrak{D}_{[A,B]}\left\{ M_{T}\right\} (\mathfrak{M})=-\frac{1}{T}\sum^{T}_{t=1}\mathfrak{M}\xi_{t}\zeta^{\top}_{t}+\zeta_{t}\xi^{\top}_{t}\mathfrak{M}^{\top}
=-\mathfrak{M}\left(\frac{1}{T}\sum^{T}_{t=1}\xi_{t}\zeta^{\top}_{t}\right)-\left(\frac{1}{T}\sum^{T}_{t=1}\zeta_{t}\xi^{\top}_{t}\right)\mathfrak{M}^{\top}.
\end{align*}
Then, putting the above and~(\ref{eq:aux-7}) into~(\ref{eq:aux-8})
we obtain 
\begin{align}
 & \mathfrak{D}_{[A,B]}l(A,B,\Sigma)(\mathfrak{M})\nonumber \\
= & -\mathrm{Tr}\left\{ \Sigma^{-1}\mathfrak{M}\left(\frac{1}{T}\sum^{T}_{t=1}\xi_{t}\zeta^{\top}_{t}\right)\right\} \nonumber \\
 & -\mathrm{Tr}\left\{ \left(\frac{1}{T}\sum^{T}_{t=1}\zeta_{t}\xi^{\top}_{t}\right)\mathfrak{M}^{\top}\Sigma^{-1}\right\} \nonumber \\
= & -2\mathrm{Tr}\left\{ \Sigma^{-1}\mathfrak{M}\left(\frac{1}{T}\sum^{T}_{t=1}\xi_{t}\zeta^{\top}_{t}\right)\right\} \nonumber \\
= & -2\mathrm{Tr}\left\{ \Sigma^{-1}\mathfrak{M}\left(\frac{1}{T}\sum^{T}_{t=1}\xi_{t}\left(x_{t+1}-\left[A,B\right]\xi_{t}\right)^{\top}\right)\right\} \nonumber \\
= & -\frac{2}{T}\mathrm{Tr}\left\{ \Sigma^{-1}\mathfrak{M}\left[V_{T}-U_{T}\left[A,B\right]^{\top}\right]\right\} .\label{eq:aux-9}
\end{align}

3) We also have, from Lemma~\ref{lem:FD-useful}-\ref{enu:cov})
\begin{equation}
\mathfrak{D}_{\Sigma}l(A,B,\Sigma)(\mathfrak{M})=\mathrm{Tr}\left\{ \Sigma^{-1}\mathfrak{M}\Sigma^{-1}\left(\Sigma-W_{T}(A,B)\right)\right\} .\label{eq:aux-10}
\end{equation}

4) The ML estimates are obtained by making 
\begin{align}
\mathfrak{D}_{[A,B]}l(A,B,\Sigma) & =0,\label{eq:aux-15}\\
\mathfrak{D}_{\Sigma}l(A,B,\Sigma) & =0.\label{eq:aux-16}
\end{align}
Since $V_{T}\in\mathrm{ran}\left(U_{T}\right)$ ($\mathrm{ran}(A)$
denotes the range of matrix $A$), in view of~(\ref{eq:aux-9}),~(\ref{eq:ML-AB})
follows from~(\ref{eq:aux-15}). Finally~(\ref{eq:ML-Sigma}) follows
immediately from~(\ref{eq:aux-10}) and~(\ref{eq:aux-16}). 
\end{IEEEproof}

\subsection{Asymptotic properties of the estimate}

\label{subsec:Asymptotic-properties-sysid}

The estimates $\hat{A}_{T}$ and$\hat{B}_{T}$ are used in place of
$A$ and $B$ to compute estimates $\hat{P}_{T}$ and $\hat{L}_{T}$
of the matrices $P$ and $L$ defining the SF-LQG controller. The
question naturally arises as to whether this is the optimal approach
for designing the controller. In this section we address this question.
More precisely, we show that $\hat{P}_{T}$ is strongly consistent
and asymptotically efficient.

Let $\vartheta=\left[\overrightarrow{A}^{\top},\overrightarrow{B}^{\top},\overset{\rightsquigarrow}{\Sigma}\right]^{\top}$
and $\mathscr{P}:\vartheta\mapsto\overset{\rightsquigarrow}{P}$ denote
the mapping induced by~(\ref{eq:Pdef}). It follows from the CRLB~\cite[eq. 3.30]{kay1993fundamentals}
that 
\begin{equation}
\mathcal{C}_{T}\geq\mathbf{J}_{\mathcal{P}}(\vartheta)\mathcal{I}^{-1}_{T}(\vartheta)\mathbf{J}^{\top}_{\mathcal{P}}(\vartheta),\label{eq:CRLB}
\end{equation}
where $\mathcal{I}_{T}(\vartheta)$ denotes the Fisher information
matrix of $X_{0,T}$ at $\vartheta$ and $\mathbf{J}_{\mathcal{P}}(\vartheta)$
denotes the Jacobian of $\mathcal{P}$ at $\vartheta$, i.e., 
\begin{align*}
\mathcal{I}_{T}(\vartheta) & =\mathcal{E}\left\{ \frac{\partial}{\partial\vartheta}\log p\left(X_{0,T};\vartheta\right)\frac{\partial}{\partial\vartheta^{\top}}\log p\left(X_{0,T};\vartheta\right)\right\} ,\\
\left[\mathbf{J}_{\mathcal{P}}(\vartheta)\right]_{i,j} & =\frac{\partial\left[\mathcal{P}\right]_{i}}{\partial\left[\vartheta\right]_{j}}(\vartheta).
\end{align*}

The following theorem states the main result of this section. 
\begin{thm}
\label{thm:ML-conv} If the random process $\left(\xi_{t}:t\in\mathbb{N}_{0}\right)$
is asymptotic mean stationary (AMS) ~\cite[S 7.3]{Gray200908} and
\begin{equation}
\liminf_{T\rightarrow\infty}\underline{\lambda}\left(\frac{1}{T}U_{T}\right)\overset{\mathrm{w.p.1}}{>}0\label{eq:assu1}
\end{equation}
($\underline{\lambda}(A)$ denotes the smallest eigenvalue of matrix
$A$), then 
\begin{align}
\lim_{T\rightarrow\infty}\hat{P}_{T}\overset{\mathrm{w.p.1}}{=} & P,\label{eq:classic-SC}\\
\lim_{T\rightarrow\infty}T\left(\mathcal{C}_{T}-\mathbf{J}_{\mathcal{P}}(\vartheta)\mathcal{I}^{-1}_{T}(\vartheta)\mathbf{J}^{\top}_{\mathcal{P}}(\vartheta)\right)= & 0.\label{eq:classic-AE}
\end{align}
\end{thm}
\begin{IEEEproof}
We split the proof in steps: 
\begin{enumerate}
\item In view of the theorem's assumption, for $\tau$ sufficiently large,
$U_{T}$ becomes non-singular for all $T\geq\tau$. We then have 
\[
\left[\hat{A}_{T},\hat{B}_{T}\right]^{\top}=U^{-1}_{T}V_{T}.
\]
Now 
\begin{align*}
V_{T} & =\sum^{T}_{t=1}\xi_{t}\left(\left[A,B\right]\xi_{t}+w_{t}\right)^{\top}\\
 & =U_{T}\left[A,B\right]^{\top}+\sum^{T}_{t=1}\xi_{t-1}w^{\top}_{t-1},
\end{align*}
Hence 
\[
\left[\hat{A}_{T},\hat{B}_{T}\right]^{\top}=\left[A,B\right]^{\top}+\left(\frac{1}{T}U_{T}\right)^{-1}\left(\frac{1}{T}\sum^{T}_{t=1}\xi_{t-1}w^{\top}_{t}\right).
\]
The AMS Ergodic theorem~\cite[Th. 8.1]{Gray200908} asserts that,
\[
\lim_{T\rightarrow\infty}\frac{1}{T}\sum^{T}_{t=1}\xi_{t-1}w^{\top}_{t}\overset{\mathrm{w.p.1}}{=}0.
\]
Hence, in view of the assumption of the theorem 
\begin{equation}
\lim_{T\rightarrow\infty}\left[\hat{A}_{T},\hat{B}_{T}\right]=\left[A,B\right].\label{eq:aux-13}
\end{equation}
It follows from~(\ref{eq:Pdef}) that $\mathscr{P}(\vartheta)$ depends
only on the components $\overrightarrow{A}$ and $\overrightarrow{B}$
of $\vartheta$. Also,~\cite[Theorem 14.2.1]{lancaster1995algebraic}
asserts that this dependence is continuous. Hence,~(\ref{eq:classic-SC})
follows from~(\ref{eq:aux-13}). 
\item It is known that under general mild regularity assumptions, the maximum
likelihood estimate is asymptotically efficient, i.e., 
\begin{equation}
\lim_{T\rightarrow\infty}T\left[\mathcal{E}\left\{ \left(\vartheta-\hat{\vartheta}_{T}\right)\left(\vartheta-\hat{\vartheta}_{T}\right)^{\top}\right\} -\mathcal{I}^{-1}_{T}(\vartheta)\right]=0,\label{eq:ML-SC}
\end{equation}
where $\hat{\vartheta}_{T}=\left[\overrightarrow{\hat{A}}^{\top}_{T},\overrightarrow{\hat{B}}^{\top}_{T},\overset{\rightsquigarrow}{\hat{\Sigma}}_{T}\right]^{\top}$.
Then, 
\begin{align*}
 & \lim_{T\rightarrow\infty}T\mathcal{C}_{T}\\
= & \lim_{T\rightarrow\infty}T\mathbf{J}_{\mathcal{P}}(\vartheta)\mathcal{E}\left\{ \left(\vartheta-\hat{\vartheta}_{T}\right)\left(\vartheta-\hat{\vartheta}_{T}\right)^{\top}\right\} \mathbf{J}^{\top}_{\mathcal{P}}(\vartheta)\\
= & \lim_{T\rightarrow\infty}T\mathbf{J}_{\mathcal{P}}(\vartheta)\mathcal{I}^{-1}_{T}(\vartheta)\mathbf{J}^{\top}_{\mathcal{P}}(\vartheta),
\end{align*}
and the result follows. 
\end{enumerate}
\end{IEEEproof}

\subsection{Complexity analysis}

\label{subsec:Complexity-analysis-sysid}

In this section we study the complexity of the algorithm consisting
of combining the system identification stage described in Section~\ref{subsec:Identification-using-maximum}
with the SF-LQG design described in Section~\ref{sec:Problem-description}.

In order to minimize the complexity of the system identification stage,
we compute~(\ref{eq:ML-AB}) using the recursive least squares (RLS)
algorithm described in~\cite[Table 13.1]{haykin2002adaptive}. This
gives 
\[
\left[\hat{A}_{\tau},\hat{B}_{\tau}\right]^{\top}=U^{-1}_{\tau}V_{\tau},
\]
where $\tau$ is the smallest integer such that $U_{\tau}$ is non-singular,
and, for all $T>\tau$, 
\begin{equation}
\left[\hat{A}_{T},\hat{B}_{T}\right]^{\top}=\left(I-\eta_{T}\xi^{\top}_{T}\right)\left[\hat{A}_{T-1},\hat{B}_{T-1}\right]^{\top}+\eta_{T}x^{\top}_{T+1},\label{eq:RLLS1}
\end{equation}
with 
\begin{align}
\eta_{T} & =\frac{M_{T-1}\xi_{T}}{1+\xi^{\top}_{T}M_{T-1}\xi_{T}},\label{eq:RLLS2}\\
M_{T} & =M_{T-1}-\eta_{T}\xi^{\top}_{T}M_{T-1}.\label{eq:RLLS3}
\end{align}
The above iterations are initialized by $M_{\tau}=U^{-1}_{\tau}$.
Also, in order to minimize the complexity of the SF-LQG stage, at
each time step $t$, the current estimate $\hat{P}_{T}$ is obtained
by updating the previous estimate $\hat{P}_{T-1}$ running one iteration
of~(\ref{eq:Pdef}). \begin{notation} We use the following: \end{notation} 
\begin{itemize}
\item $\mathrm{GAUSS}(N)$ denotes the complexity of solving a system of
$N$ linear equations with $N$ unknowns using Gaussian elimination.
It is given by 
\[
\mathrm{GAUSS}(N)=\frac{4N^{3}+9N^{2}-5N}{6}.
\]
\item $\mathrm{DARE}(N,M)$ denotes the complexity of computing a single
iteration of~(\ref{eq:Pdef}) to update the current approximate solution
of a DARE with $N$ states and $M$ inputs. It is given by 
\begin{align*}
\mathrm{DARE}(N,M) & =2N^{3}+N^{2}+3N^{2}M+NM^{2}+M^{2}\\
 & +N\times\mathrm{GAUSS}(M).
\end{align*}
\item $\mathrm{RLLS}(N)$ denotes the complexity of running a single iteration
of the recursive LLS algorithm~(\ref{eq:RLLS1})-(\ref{eq:RLLS3}),
for solving a problem with an $N\times M$ unknown matrix. It is given
by 
\[
\mathrm{RLLS}(N,M)=NM(3N+4).
\]
\end{itemize}
Using the above notation, the complexity per sample time of the algorithm
is given by: 
\begin{enumerate}
\item $\mathrm{RLLS}(N+M,N)$ for running one iteration of~(\ref{eq:RLLS1}). 
\item $\mathrm{DARE}(N,M)$ for computing the right-hand side of~(\ref{eq:Pdef}). 
\item $2N^{2}M+NM\left(1+M\right)+N\times\mathrm{GAUSS}(M)$ to compute~(\ref{eq:Ldef}). 
\end{enumerate}

\section{Solution using Q-learning}

\label{sec:Solution-using-Q-learning}

In this Section we study the alternative approach for solving the
problem, which consists in using the Q-learning algorithm. In Section~\ref{subsec:Proposed-method}
we propose a novel approach for doing so. As mentioned, an advantage
of this approach is that it permits guaranteeing the strong consistency
of the estimate $\hat{P}_{T}$. In Section~\ref{subsec:Asymptotic-properties-qlearning}
we show this result. Finally, in Section~\ref{subsec:Complexity-analysis-qlearning}
we propose a numerically efficient implementation of the proposed
algorithm and compute its complexity.

\subsection{Proposed method}

\label{subsec:Proposed-method}

Let 
\begin{equation}
\Lambda=\left[\begin{array}{cc}
\Lambda_{1,1} & \Lambda_{1,2}\\
\Lambda_{2,1} & \Lambda_{2,2}
\end{array}\right],\label{eq:Lambda-def}
\end{equation}
with 
\begin{align*}
\Lambda_{1,1} & =\gamma A^{\top}PA+Q, & \Lambda_{1,2} & =\gamma A^{\top}PB,\\
\Lambda_{2,1} & =\gamma B^{\top}PA, & \Lambda_{2,2} & =\gamma B^{\top}PB+R.
\end{align*}
We can parameterize $L$ and $P$, defined in~(\ref{eq:Ldef}) and~(\ref{eq:Pdef}),
respectively, in terms of $\Lambda$. We do this as follows 
\begin{align}
L(\Lambda) & =\Lambda^{-1}_{2,2}\Lambda_{2,1},\label{eq:L(Lambda)}\\
P(\Lambda) & =\Lambda_{1,1}-\Lambda_{1,2}\Lambda^{-1}_{2,2}\Lambda_{2,1}.\label{eq:S(Lambda)}
\end{align}
Hence, we can address our research problem by estimating $\Lambda$.

Towards this end, we start by defining the following Q-function 
\begin{align}
\mathcal{Q}\left(x_{t},u_{t}\right) & =x^{\top}_{t}Qx_{t}+u^{\top}_{t}Ru_{t}\nonumber \\
 & +\gamma\mathcal{E}\left\{ \min_{u_{t+1}}\mathcal{Q}\left(x_{t+1},u_{t+1}\right)\,\middle|\,x_{t}\right\} \nonumber \\
 & =\xi^{\top}_{t}\Delta\xi_{t}+\gamma\mathcal{E}\left\{ \min_{u_{t+1}}\mathcal{Q}\left(x_{t+1},u_{t+1}\right)\,\middle|\,x_{t}\right\} ,\label{eq:Qfun-def}
\end{align}
where $\Delta=\mathrm{diag}\left(Q,R\right)$. It then follows that
\[
J_{t}\left(x_{t},U^{\star}_{t,\infty}\left(x_{t}\right)\right)=\min_{u_{1}}\mathcal{Q}\left(x_{1},u_{1}\right).
\]

The Q-learning method consists in obtaining an estimate of (i.e.,
learning) the Q-function, at time $t$, using the data $U_{0,t-t}$
and $X_{0,t}$ available up to that time. In our problem, the state
and input spaces are uncountable sets. This makes it impossible to
directly apply the Q-learning formulas to address our problem. To
tackle this, we would like to parameterize the Q-function $\mathcal{Q}$
in terms of $\Lambda$. It turns out that this is not readily possible.
But as pointed out in~\cite{kumar2023unconstrained}, it is possible
to parameterize $\mathcal{Q}$ in terms of $\theta^{\top}=\left[\overset{\rightsquigarrow}{\Lambda}^{\top},\eta\right]$,
where $\eta=\gamma\mathrm{Tr}\left\{ \Sigma P(\Lambda)\right\} $.
This is stated in the next lemma, where we use $Q\left(x_{t},u_{t};\theta\right)$
to denote the parametric version of $\mathcal{Q}\left(x_{t},u_{t}\right)$. 
\begin{lem}
\label{lem:parametric-Q} The following holds true 
\begin{equation}
\mathcal{Q}\left(x_{t},u_{t};\theta\right)=\xi^{\top}_{t}\Lambda\xi_{t}+\frac{\eta}{1-\gamma}.\label{eq:Qfun-exp}
\end{equation}
\end{lem}
\begin{IEEEproof}
Form the classic solution of the SF-LQG problem we obtain 
\begin{align*}
 & \mathcal{E}\left\{ \min_{u_{t+1}}\mathcal{Q}\left(x_{t+1},u_{t+1}\right)|x_{t}\right\} \\
= & \mathcal{E}\left\{ J\left(x_{t+1},U^{\star}\left(x_{t+1}\right)\right)|x_{t}\right\} \\
= & \mathcal{E}\left\{ x^{\top}_{t+1}P(\Lambda)x_{t+1}|x_{t}\right\} +\frac{\gamma}{1-\gamma}\mathrm{Tr}\left\{ \Sigma P(\Lambda)\right\} \\
= & \mathcal{E}\left\{ \left(Ax_{t}+Bu_{t}+w_{t}\right)^{\top}P(\Lambda)\left(Ax_{t}+Bu_{t}+w_{t}\right)|x_{t}\right\} \\
 & +\frac{\gamma}{1-\gamma}\mathrm{Tr}\left\{ \Sigma P(\Lambda)\right\} \\
= & \left(Ax_{t}+Bu_{t}\right)^{\top}P(\Lambda)\left(Ax_{t}+Bu_{t}\right)+\frac{1}{1-\gamma}\mathrm{Tr}\left\{ \Sigma P(\Lambda)\right\} \\
= & \xi^{\top}_{t}\left[\begin{array}{c}
A^{\top}\\
B^{\top}
\end{array}\right]P(\Lambda)\left[\begin{array}{cc}
A & B\end{array}\right]\xi_{t}+\frac{1}{1-\gamma}\mathrm{Tr}\left\{ \Sigma P(\Lambda)\right\} .
\end{align*}
Putting the above into~(\ref{eq:Qfun-def}), and using~(\ref{eq:Lambda-def})
we get 
\begin{align*}
\mathcal{Q}\left(x_{t},u_{t}\right) & =\xi^{\top}_{t}\left\{ \Delta+\gamma\left[\begin{array}{c}
A^{\top}\\
B^{\top}
\end{array}\right]P(\Lambda)\left[\begin{array}{cc}
A & B\end{array}\right]\right\} \xi_{t}\\
 & +\frac{\gamma}{1-\gamma}\mathrm{Tr}\left\{ \Sigma P(\Lambda)\right\} \\
 & =\xi^{\top}_{t}\Lambda\xi_{t}+\frac{\gamma}{1-\gamma}\mathrm{Tr}\left\{ \Sigma P(\Lambda)\right\} .
\end{align*}
and the result follows. 
\end{IEEEproof}
In view of Lemma~\ref{lem:parametric-Q}, the problem of estimating
$\mathcal{Q}$ turns into that of estimating $\theta$. The next lemma
gives us the first step towards addressing this problem. 
\begin{lem}
\label{lem:criterion} For all $t\in\llbracket T\rrbracket$, 
\begin{equation}
\phi^{\top}_{t}\theta=\xi^{\top}_{t}\Delta\xi_{t}+\gamma x^{\top}_{t+1}P(\Lambda)x_{t+1}-\gamma\epsilon_{t},\label{eq:est-crit-2}
\end{equation}
where $\phi^{\top}_{t}=\left[\left(\xi_{t}\tilde{\otimes}\xi_{t}\right)^{\top},1\right]$
and 
\begin{equation}
\epsilon_{t}=x^{\top}_{t+1}P(\Lambda)x_{t+1}-\mathcal{E}\left\{ x^{\top}_{t+1}P(\Lambda)x_{t+1}\left|x_{t}\right.\right\} .\label{eq:N}
\end{equation}
\end{lem}
\begin{IEEEproof}
From~(\ref{eq:Qfun-def}) and~(\ref{eq:Qfun-exp}) 
\begin{align*}
 & \xi^{\top}_{t}\Lambda\xi_{t}+\frac{\eta}{1-\gamma}\\
= & \xi^{\top}_{t}\Delta\xi_{t}+\gamma\mathcal{E}\left\{ \min_{u_{t+1}}\xi^{\top}_{t+1}\Lambda\xi_{t+1}+\frac{\eta}{1-\gamma}\left|x_{t}\right.\right\} \\
= & \xi^{\top}_{t}\Delta\xi_{t}+\gamma\mathcal{E}\left\{ \min_{u_{t+1}}\xi^{\top}_{t+1}\Lambda\xi_{t+1}\left|x_{t}\right.\right\} +\frac{\gamma\eta}{1-\gamma}\\
= & \xi^{\top}_{t}\Delta\xi_{t}+\gamma\mathcal{E}\left\{ \xi^{\star\top}_{t+1}(\Lambda)\Lambda\xi^{\star}_{t+1}(\Lambda)\left|x_{t}\right.\right\} +\frac{\gamma\eta}{1-\gamma},
\end{align*}
where $\xi^{\star\top}_{t}(\Lambda)=\left[x^{\top}_{t},\left(-L(\Lambda)x_{t}\right)^{\top}\right]$.
Then 
\begin{align}
\xi^{\top}_{t}\Lambda\xi_{t} & =\xi^{\top}_{t}\Delta\xi_{t}+\gamma\mathcal{E}\left\{ \xi^{\star\top}_{t+1}(\Lambda)\Lambda\xi^{\star}_{t+1}(\Lambda)\left|x_{t}\right.\right\} -\eta\nonumber \\
 & =\xi^{\top}_{t}\Delta\xi_{t}+\gamma\mathcal{E}\left\{ x^{\top}_{t+1}P(\Lambda)x_{t+1}\left|x_{t}\right.\right\} -\eta.\label{eq:est-crit-1}
\end{align}
Using~(\ref{eq:N}) in~(\ref{eq:est-crit-1}) we obtain 
\[
\xi^{\top}_{t}\Lambda\xi_{t}=\xi^{\top}_{t}\Delta\xi_{t}+\gamma x^{\top}_{t+1}P(\Lambda)x_{t+1}-\eta-\gamma\epsilon_{t}.
\]
We can rewrite the above as 
\begin{align*}
\left[\left(\xi_{t}\tilde{\otimes}\xi_{t}\right)^{\top},1\right]\left[\begin{array}{c}
\overset{\rightsquigarrow}{\Lambda}\\
\eta
\end{array}\right]
=\xi^{\top}_{t}\Delta\xi_{t}+\gamma x^{\top}_{t+1}P(\Lambda)x_{t+1}-\gamma\epsilon_{t},
\end{align*}
giving the result. 
\end{IEEEproof}
Lemma~\ref{lem:criterion} states that we can estimate $\theta$
by iterating equation~(\ref{eq:est-crit-2}), in which the term $\epsilon_{t}$
acts as additive noise. The following lemma gives the statistical
properties of this noise term. 
\begin{lem}
\label{lem:noise} Conditioned on $\xi_{t}$, the noise term $\epsilon_{t}$
has generalized chi-squared distribution (see Appendix~\ref{sec:Generalized-chi-squared-distribu})
given by 
\begin{align}
\epsilon_{t}\sim\bar{\chi}\left(\Sigma^{1/2}P(\Lambda)\Sigma^{1/2},\right.
\left.2\Sigma^{1/2}P(\Lambda)\left(Ax_{t}+Bu_{t}\right),-\mathrm{Tr}\left\{ \Sigma P(\Lambda)\right\} \right).\label{eq:N-dist}
\end{align}
Also 
\begin{align*}
\mathcal{E}\left\{ \epsilon_{t}\right\}  & =0,\\
\mathcal{V}\left\{ \epsilon_{t}\right\}  & =2\mathrm{Tr}\left\{ P(\Lambda)\Sigma P(\Lambda)\Sigma\right\} \\
 & +2\xi^{\top}_{t}\left[\begin{array}{c}
A^{\top}\\
B^{\top}
\end{array}\right]P(\Lambda)\Sigma P(\Lambda)\left[\begin{array}{cc}
A & B\end{array}\right]\xi_{t}.
\end{align*}
\end{lem}
\begin{IEEEproof}
We have 
\begin{align}
 & x^{\top}_{t+1}P(\Lambda)x_{t+1}\nonumber \\
= & \left(Ax_{t}+Bu_{t}+w_{t}\right)^{\top}P(\Lambda)\left(Ax_{t}+Bu_{t}+w_{t}\right)\nonumber \\
= & w^{\top}_{t}P(\Lambda)w_{t}+2\left(Ax_{t}+Bu_{t}\right)^{\top}P(\Lambda)w_{t}\nonumber \\
 & +\left(Ax_{t}+Bu_{t}\right)^{\top}P(\Lambda)\left(Ax_{t}+Bu_{t}\right),\label{eq:N1}
\end{align}
and 
\begin{align}
\mathcal{E}\left\{ x^{\top}_{t+1}P(\Lambda)x_{t+1}\left|x_{t}\right.\right\} =\mathrm{Tr}\left\{ P(\Lambda)\Sigma\right\} 
+\left(Ax_{t}+Bu_{t}\right)^{\top}P(\Lambda)\left(Ax_{t}+Bu_{t}\right).\label{eq:N2}
\end{align}
Putting~(\ref{eq:N1})-(\ref{eq:N2}) into~(\ref{eq:N}) we obtain
\begin{align}
\epsilon_{t}=w^{\top}_{t}P(\Lambda)w_{t}
+2\left(Ax_{t}+Bu_{t}\right)^{\top}P(\Lambda)w_{t}-\mathrm{Tr}\left\{ P(\Lambda)\Sigma\right\} .\label{eq:epsilon_t}
\end{align}
The first follows from the above equation and Appendix~\ref{sec:Generalized-chi-squared-distribu}.
Also, from Proposition~\ref{prop:GCS-mean-var}

\begin{align}
 & \mathcal{E}\left\{ \epsilon_{t}\right\} \nonumber \\
= & \mathrm{Tr}\left\{ \Sigma^{1/2}P(\Lambda)\Sigma^{1/2}\right\} -\mathrm{Tr}\left\{ P(\Lambda)\Sigma\right\} \nonumber \\
= & 0,\label{eq:N-mean}
\end{align}
and 
\begin{align*}
\mathcal{V}\left\{ \epsilon_{t}\right\}  & =2\mathrm{Tr}\left\{ \Sigma^{1/2}P(\Lambda)\Sigma P(\Lambda)\Sigma^{1/2}\right\} \\
 & +2\left(Ax_{t}+Bu_{t}\right)^{\top}P(\Lambda)\Sigma P(\Lambda)\left(Ax_{t}+Bu_{t}\right),
\end{align*}
and the result follows. 
\end{IEEEproof}
In view of Lemmas~\ref{lem:criterion} and~\ref{lem:noise}, we
can derive a method for estimating $\theta$. We start by writing
equation~(\ref{eq:est-crit-2}) as follows 
\begin{equation}
\Upsilon_{T}\left(\theta\right)=\Phi^{\top}_{T}\theta+E_{T},\label{eq:Lambda-est-1}
\end{equation}
where $\Phi_{T}=\left[\phi_{1},\cdots,\phi_{T}\right]$. Also, $\Upsilon^{\top}_{T}\left(\theta\right)=\left[\upsilon_{1}\left(\theta\right),\cdots,\upsilon_{T}\left(\theta\right)\right]$
and $E^{\top}_{T}=\left[e_{1},\cdots,e_{T}\right]$ are given by 
\begin{align*}
\upsilon_{t}\left(\theta\right) & =\xi^{\top}_{t}\Delta\xi_{t}+\gamma x^{\top}_{t+1}P(\Lambda)x_{t+1},\\
e_{t} & =\gamma\epsilon_{t}.
\end{align*}
Using~(\ref{eq:Lambda-est-1}) we can infer the following recursive
method for obtaining, at time $T$, an estimate of $\theta$ 
\begin{align}
\theta^{k+1}_{T} & =\left(\Phi_{T}\Phi^{\top}_{T}\right)^{-1}\Phi_{T}\Upsilon_{T}\left(\theta^{k}_{T}\right)\nonumber \\
 & =U^{-1}_{T}V_{T}\left(\theta^{k}_{T}\right),\label{eq:algorithm}
\end{align}
where 
\begin{align*}
U_{T} & =\sum^{T}_{t=1}\phi_{t}\phi^{\top}_{t},\\
V_{T}(\theta) & =\sum^{T}_{t=1}\phi_{t}\upsilon_{t}(\theta).
\end{align*}
Notice that in the above recursions the value of $T$ is fixed. Since
we aim for an online algorithm, we make $T=k$, i.e., we do one iteration
of~(\ref{eq:Lambda-est-1}) for every sample time $T$. Defining
$\theta_{T}=\theta^{T}_{T}$, this leads to the following recursions

\begin{equation}
\theta_{T}=U^{-1}_{T}V_{T}\left(\theta_{T-1}\right).\label{eq:online-algorithm}
\end{equation}

\begin{rem}
Notice that, to implement~(\ref{eq:online-algorithm}), we in principle
need to recompute, at each time step $T$, the whole history $\upsilon_{1}\left(\theta\right),\cdots,\upsilon_{T}\left(\theta\right)$.
Doing so would lead to an algorithm whose complexity increases with
$T$. In Section~\ref{subsec:Complexity-analysis-qlearning} we describe
a numerically efficient implementation of~(\ref{eq:online-algorithm})
which avoids this problem. 
\end{rem}

\subsection{Asymptotic properties of the estimate}

\label{subsec:Asymptotic-properties-qlearning}

In this section we show that the recursions~(\ref{eq:online-algorithm})
lead to a strongly consistent estimate $\hat{P}_{T}$. This is stated
in the next result. 
\begin{thm}
\label{thm:qlearning-conv} If the random process $\left(\xi_{t}:t\in\mathbb{N}_{0}\right)$
is asymptotic mean stationary~\cite[S 7.3]{Gray200908} and 
\begin{equation}
\liminf_{T\rightarrow\infty}\underline{\lambda}\left(\frac{1}{T}U_{T}\right)\overset{\mathrm{w.p.1}}{>}0.\label{eq:assu2}
\end{equation}
then 
\[
\lim_{t\rightarrow\infty}\theta_{t}\overset{\mathrm{w.p.1}}{=}\theta.
\]
\end{thm}
\begin{IEEEproof}
We split the proof in steps:

1) Let $\bar{\Upsilon}^{\top}_{T}(\theta)=\left[\bar{\upsilon}_{1}(\theta),\cdots,\bar{\upsilon}_{T}(\theta)\right]$
and $\tilde{\Upsilon}^{\top}_{T}(\theta)=\left[\tilde{\upsilon}_{1}(\theta),\cdots,\tilde{\upsilon}_{T}(\theta)\right]$
be defined by 
\begin{align*}
\bar{\upsilon}_{t}\left(\theta\right) & =\xi^{\top}_{t}\Delta\xi_{t}+\gamma\mathcal{E}\left\{ x^{\top}_{t+1}P\left(\Lambda\right)x_{t+1}\left|x_{t}\right.\right\} ,\\
\tilde{\upsilon}_{t}\left(\theta\right) & =\gamma\epsilon_{t}.
\end{align*}
It then follows that 
\[
\Upsilon_{T}\left(\theta^{k}_{T}\right)_{t}=\bar{\Upsilon}_{T}\left(\theta^{k}_{T}\right)+\tilde{\Upsilon}_{T}\left(\theta^{k}_{T}\right).
\]
We can then write~(\ref{eq:algorithm}) as 
\begin{equation}
\theta^{k+1}_{T}=\bar{\theta}^{k+1}_{T}+\tilde{\theta}^{k+1}_{T},\label{eq:theta-decomposition}
\end{equation}
with 
\begin{align}
\bar{\theta}^{k+1}_{T} & =\left(\Phi_{T}\Phi^{\top}_{T}\right)^{-1}\Phi_{T}\bar{\Upsilon}_{T}\left(\theta^{k}_{T}\right),\label{eq:theta-bar}\\
\tilde{\theta}^{k+1}_{T} & =\left(\Phi_{T}\Phi^{\top}_{T}\right)^{-1}\Phi_{T}\tilde{\Upsilon}_{T}\left(\theta^{k}_{T}\right).\label{eq:theta-tilde}
\end{align}

2) We have 
\begin{equation}
\bar{\theta}^{k+1}_{T}\in\underset{\vartheta}{\arg\min}\left\Vert \Phi^{\top}_{T}\vartheta-\bar{\Upsilon}_{T}\left(\theta^{k}_{T}\right)\right\Vert .\label{eq:LLS}
\end{equation}
It follows from~(\ref{eq:assu2}) that, for $T$ sufficiently large,
$\Phi_{T}$ has full row rank. Therefore the above has a unique solution.
Now, from the definition of $\theta$, 
\begin{equation}
\phi^{\top}_{t}\bar{\theta}^{k+1}_{T}=\xi^{\top}_{t}\bar{\Lambda}^{k+1}_{T}\xi_{t}+\bar{\eta}^{k+1}_{T}.\label{eq:aux-1}
\end{equation}
Also 
\begin{equation}
\begin{aligned}
\bar{\Upsilon}_{t}\left(\theta^{k}_{T}\right)=&\xi^{\top}_{t}\Delta\xi_{t}+\gamma\mathcal{E}\left\{ x^{\top}_{t+1}P\left(\Lambda^{k}_{T}\right)x_{t+1}\left|x_{t}\right.\right\} \\
=&\xi^{\top}_{t}\Delta\xi_{t}+\gamma\xi^{\top}_{t}\left[\begin{array}{c}
A^{\top}\\
B^{\top}
\end{array}\right]P\left(\Lambda^{k}\right)\left[\begin{array}{cc}
A & B\end{array}\right]\xi_{t}
+\gamma\mathcal{E}\left\{ w^{\top}_{t}P\left(\Lambda^{k}\right)w_{t}\right\} \\
=&\xi^{\top}_{t}\left\{ \Delta+\gamma\left[\begin{array}{c}
A^{\top}\\
B^{\top}
\end{array}\right]P\left(\Lambda^{k}_{T}\right)\left[\begin{array}{cc}
A & B\end{array}\right]\right\} \xi_{t}
+\gamma\mathrm{Tr}\left\{ \Sigma P\left(\Lambda^{k}_{T}\right)\right\} .\label{eq:aux-2}
\end{aligned}
\end{equation}
From~(\ref{eq:aux-1})-(\ref{eq:aux-2}), for all $t\in\llbracket T\rrbracket$,
\begin{align*}
\phi^{\top}_{t}\bar{\theta}^{k+1}_{T}-\bar{\Upsilon}_{t}\left(\theta^{k}_{T}\right)
=\xi^{\top}_{t}\left\{ \bar{\Lambda}^{k+1}_{T}-\Delta-\gamma\left[\begin{array}{c}
A^{\top}\\
B^{\top}
\end{array}\right]P\left(\Lambda^{k}_{T}\right)\left[\begin{array}{cc}
A & B\end{array}\right]\right\} \xi_{t}
+\bar{\eta}^{k+1}_{T}-\gamma\mathrm{Tr}\left\{ \Sigma P\left(\Lambda^{k}_{T}\right)\right\} .
\end{align*}
Then, since the solution of~(\ref{eq:LLS}) is unique, it is given
by 
\begin{align}
\bar{\Lambda}^{k+1}_{T} & =\Delta+\gamma\left[\begin{array}{c}
A^{\top}\\
B^{\top}
\end{array}\right]P\left(\Lambda^{k}_{T}\right)\left[\begin{array}{cc}
A & B\end{array}\right],\label{eq:Lambda-it-noiseless}\\
\bar{\eta}^{k+1}_{T} & =\gamma\mathrm{Tr}\left\{ \Sigma P\left(\Lambda^{k}_{T}\right)\right\} .\label{eq:eta-it-noiseless}
\end{align}

3) From~(\ref{eq:theta-tilde}) we have 
\begin{equation}
\tilde{\theta}^{k+1}_{T}=\left(\frac{1}{T}U_{T}\right)^{-1}\frac{\gamma}{T}\sum^{T}_{t=1}\phi_{t}\epsilon_{t}.\label{eq:aux-3}
\end{equation}
Now, from~(\ref{eq:epsilon_t}), 
\begin{align*}
\phi_{t}\epsilon_{t} & =\phi_{t}w^{\top}_{t}P(\Lambda)w_{t}+2\phi_{t}\left(Ax_{t}+Bu_{t}\right)^{\top}P(\Lambda)w_{t}\\
 & -\phi_{t}\mathrm{Tr}\left\{ P(\Lambda)\Sigma\right\} ,
\end{align*}
from where immediately follows that 
\[
\mathcal{E}\left\{ \phi_{t}\epsilon_{t}\right\} =0.
\]
Then, the AMS ergodic theorem~~\cite[Th. 8.1]{Gray200908} asserts
that 
\[
\lim_{T\rightarrow\infty}\frac{1}{T}\sum^{T}_{t=1}\phi_{t}\epsilon_{t}\overset{\mathrm{w.p.1}}{=}\lim_{T\rightarrow\infty}\frac{1}{T}\sum^{T}_{t=1}\mathcal{E}\left\{ \phi_{t}\epsilon_{t}\right\} =0.
\]
It then follows from the assumption of the theorem and~(\ref{eq:aux-3})
that 
\[
\lim_{T\rightarrow\infty}\tilde{\theta}^{k+1}_{T}\overset{\mathrm{w.p.1}}{=}0.
\]

4) From~(\ref{eq:theta-decomposition}),~(\ref{eq:Lambda-it-noiseless})
and~(\ref{eq:eta-it-noiseless}) we have 
\begin{align}
\Lambda^{k+1}_{T} & =\Delta+\gamma\left[\begin{array}{c}
A^{\top}\\
B^{\top}
\end{array}\right]P\left(\Lambda^{k}_{T}\right)\left[\begin{array}{cc}
A & B\end{array}\right]\nonumber \\
 & +\left[\begin{array}{cc}
E^{(1,1)}_{T} & E^{(1,2)}_{T}\\
E^{(2,1)}_{T} & E^{(2,2)}_{T}
\end{array}\right],\label{eq:Lambda-full}\\
\eta^{k+1}_{T} & =\gamma\mathrm{Tr}\left\{ \Sigma P\left(\Lambda^{k}_{T}\right)\right\} +e_{T}.\label{eq:eta-full}
\end{align}
with 
\begin{eqnarray}
\lim_{T\rightarrow\infty}E^{(i,j)}_{T} & \overset{\mathrm{w.p.1}}{=} & 0,\;\forall i,j\in\{1,2\},\label{eq:aux-4}\\
\lim_{T\rightarrow\infty}e_{T} & \overset{\mathrm{w.p.1}}{=} & 0.\label{eq:aux-12}
\end{eqnarray}

5) From~(\ref{eq:Lambda-full}) we have 
\begin{align*}
P\left(\Lambda^{k+1}_{T}\right)= & \left[\Lambda^{k+1}_{T}\right]_{11}-\left[\Lambda^{k+1}_{T}\right]_{12}\left[\Lambda^{k+1}_{T}\right]^{-1}_{22}\left[\Lambda^{k+1}_{T}\right]_{21}\\
= & \left[\gamma A^{\top}P\left(\Lambda^{k}_{T}\right)A+Q+E^{(1,1)}_{T}\right]-\\
 & \left[\gamma A^{\top}P\left(\Lambda^{k}_{T}\right)B+E^{(1,2)}_{T}\right]\times\\
 & \times\left[\gamma B^{\top}P\left(\Lambda^{k}_{T}\right)B+R+E^{(2,2)}_{T}\right]^{-1}\times\\
 & \times\left[\gamma B^{\top}P\left(\Lambda^{k}_{T}\right)A+E^{(2,1)}_{T}\right].
\end{align*}
Letting $P^{k}_{T}=P\left(\Lambda^{k}_{T}\right)$ we get 
\begin{align*}
P^{k+1}_{T} & =\left[\tilde{A}^{\top}P^{k}_{T}\tilde{A}+Q+E^{(1,1)}_{T}\right]-\left[\tilde{A}^{\top}P^{k}_{T}\tilde{B}+E^{(1,2)}_{T}\right]\times\\
 & \times\left[\tilde{B}^{\top}P^{k}_{T}\tilde{B}+R+E^{(2,2)}_{T}\right]^{-1}\left[\tilde{B}^{\top}P^{k}_{T}\tilde{A}+E^{(2,1)}_{T}\right],
\end{align*}
where $\tilde{A}=\sqrt{\gamma}A$ and $\tilde{B}=\sqrt{\gamma}B$
. In view of~\cite[Proposition 12.1.1]{lancaster1995algebraic},
the above is in turn equal to 
\begin{align*}
P^{k+1}_{T} & =\breve{A}^{\top}_{T}P^{k}_{T}\breve{A}_{T}+\breve{Q}_{T}-\breve{A}^{\top}_{T}P^{k}_{T}\tilde{B}\left[\breve{R}_{T}+\tilde{B}^{\top}P^{k}_{T}\tilde{B}\right]^{-1}\times\\
 & \times\tilde{B}^{\top}P^{k}_{T}\breve{A}_{T},
\end{align*}
where 
\begin{align*}
\breve{A}_{T} & =\tilde{A}-\sqrt{\gamma}B\left(R+E^{(2,2)}_{T}\right)^{-1}E^{(2,1)}_{T},\\
\breve{Q}_{T} & =Q+E^{(1,1)}_{T}-E^{(1,2)}_{T}\left(R+E^{(2,2)}_{T}\right)^{-1}E^{(2,1)}_{T},\\
\breve{R}_{T} & =R+E^{(2,2)}_{T}.
\end{align*}

6) It follows from Proposition~\ref{pro:DARE} that there exists
a unique $P>0$ satisfying 
\[
P=\tilde{A}^{\top}P\tilde{A}^{\top}+Q-\tilde{A}^{\top}P\tilde{B}\left[R+\tilde{B}^{\top}P\tilde{B}\right]^{-1}\tilde{B}^{\top}P\tilde{A}.
\]
From~(\ref{eq:aux-4}), 
\begin{align*}
\lim_{T\rightarrow\infty}\breve{A}_{T} & \overset{\mathrm{w.p.1}}{=}\tilde{A}, & \lim_{T\rightarrow\infty}\breve{Q}_{T} & \overset{\mathrm{w.p.1}}{=}Q,\\
\lim_{T\rightarrow\infty}\breve{R}_{T} & \overset{\mathrm{w.p.1}}{=}R.
\end{align*}
Then, w.p.1, there exists $\underline{T}$ such that for all $T\geq\underline{T}$,
the conditions of Proposition~\ref{pro:DARE} hold. Hence, for each
$T\geq\underline{T}$, there exists a unique $P_{T}>0$ satisfying
\begin{align*}
P_{T} & =\breve{A}^{\top}_{T}P_{T}\breve{A}_{T}+\breve{Q}_{T}-\breve{A}^{\top}_{T}P_{T}\tilde{B}\times\\
 & \times\left[\breve{R}_{T}+\tilde{B}^{\top}P_{T}\tilde{B}\right]^{-1}\tilde{B}^{\top}P_{T}\breve{A}_{T}.
\end{align*}
Then, from~\cite[Theorem 14.2.1]{lancaster1995algebraic} 
\[
\lim_{T\rightarrow\infty}P_{T}\overset{\mathrm{w.p.1}}{=}P.
\]

7) Let 
\[
f_{k}(T)=\left\Vert P^{k}_{T}-P_{T}\right\Vert .
\]
From Proposition~\ref{pro:DARE}, for all $T\in\mathbb{N}$, 
\begin{align*}
f_{k}(T) & \leq\left(e^{\rho^{k}\delta\left(P_{T},P^{0}_{0}\right)}-1\right)\left\Vert P_{T}\right\Vert \\
 & \leq a\left(e^{\rho^{k}b}-1\right),
\end{align*}
where 
\begin{align*}
a & =\sup_{T\in\mathbb{N}}\left\Vert P_{T}\right\Vert <\infty,\\
b & =\sup_{T\in\mathbb{N}}\delta\left(P_{T},P^{0}_{0}\right)<\infty.
\end{align*}
This means that $f_{k}$ tends to $0$ uniformly. Thus $P^{k}_{T}$
tends to $P_{T}$ uniformly. It then follows that 
\[
\lim_{T\rightarrow\infty}P^{T}_{T}=P.
\]

8) Since $\Lambda$ is determined by $P$, we have 
\begin{equation}
\lim_{T\rightarrow\infty}\Lambda^{T}_{T}=\Lambda.\label{eq:lim-Lambda}
\end{equation}
Also, from~(\ref{eq:eta-full}) and~(\ref{eq:aux-12}) 
\begin{equation}
\lim_{T\rightarrow\infty}\eta^{T}_{T}=\gamma\mathrm{Tr}\left\{ \Sigma P\left(\Lambda\right)\right\} .\label{eq:lim-eta}
\end{equation}
The result then follows from~(\ref{eq:lim-Lambda}) and~(\ref{eq:lim-eta}). 
\end{IEEEproof}

\subsection{Complexity analysis}

\label{subsec:Complexity-analysis-qlearning}

In this section we study the complexity of the algorithm introduced
in Section~\ref{subsec:Proposed-method}. We start by deriving a
recursive implementation of~(\ref{eq:online-algorithm}). This implementation
is in the line of the RLS algorithm described in Section~\ref{subsec:Complexity-analysis-sysid}. 
\begin{prop}
\label{prop:RLS-qlearning}Let $\tau$ be the smallest integer such
that $U_{\tau}$ is non-singular. Then, for all $T>\tau,$ 
\[
\theta_{T}=\vartheta_{T}+\Theta_{T}\overset{\rightsquigarrow}{P}\left(\Lambda_{T-1}\right),
\]
where 
\begin{align*}
\vartheta_{T} & =\left(I-\eta_{T}\phi^{\top}_{T}\right)\vartheta_{T-1}+\eta_{T}\xi^{\top}_{T}\Delta\xi_{T},\\
\Theta_{T} & =\left(I-\eta_{T}\phi^{\top}_{T}\right)\Theta_{T-1}+\gamma\eta_{T}\left(x_{T+1}\tilde{\otimes}x_{T+1}\right)^{\top},
\end{align*}
and 
\begin{align*}
\eta_{T} & =\frac{M_{T-1}\phi_{T}}{1-\phi^{\top}_{T}M_{T-1}\phi_{T}},\\
M_{T} & =M_{T-1}-\eta_{T}\phi^{\top}_{T}M_{T-1}.
\end{align*}
The above iterations are initialized by $M_{\tau}=U^{-1}_{\tau}$,
and 
\begin{align*}
\vartheta_{\tau} & =U^{-1}_{\tau}\sum^{\tau}_{t=1}\phi_{t}\xi^{\top}_{t}\Delta\xi_{t},\\
\Theta_{\tau} & =\gamma U^{-1}_{\tau}\sum^{\tau}_{t=1}\phi_{t}\left(x_{t+1}\otimes x_{t+1}\right)^{\top}.
\end{align*}
\end{prop}
\begin{IEEEproof}
We split the proof in steps:

1) Let 
\begin{align*}
U_{T} & =\Phi_{T}\Phi^{\top}_{T},\\
V_{T}(\theta) & =\Phi_{T}\Upsilon_{T}(\theta).
\end{align*}
Then 
\[
U_{T}=\phi_{T}\phi^{\top}_{T}+U_{T-1},
\]
and 
\begin{align*}
V_{T}(\theta) & =\sum^{T}_{t=1}\phi_{t}\upsilon_{t}(\theta)\\
 & =\sum^{T}_{t=1}\phi_{t}\xi^{\top}_{t}\Delta\xi_{t}+\gamma\sum^{T}_{t=1}\phi_{t}x^{\top}_{t+1}P(\Lambda)x_{t+1}\\
 & =\sum^{T}_{t=1}\phi_{t}\xi^{\top}_{t}\Delta\xi_{t}+\gamma\sum^{T}_{t=1}\phi_{t}\left(x_{t+1}\tilde{\otimes}x_{t+1}\right)^{\top}\overset{\rightsquigarrow}{P}(\Lambda)\\
 & =a_{T}+B_{T}\overset{\rightsquigarrow}{P}(\Lambda).
\end{align*}
where 
\begin{align*}
a_{T} & =\sum^{T}_{t=1}\phi_{t}\xi^{\top}_{t}\Delta\xi_{t},\\
B_{T} & =\gamma\sum^{T}_{t=1}\phi_{t}\left(x_{t+1}\otimes x_{t+1}\right)^{\top}.
\end{align*}
Letting $M_{T}=U^{-1}_{T}$ it follows that 
\begin{align*}
\theta_{T} & =U^{-1}_{T}V_{T}(\theta)\\
 & =M_{T}V_{T}(\theta)\\
 & =M_{T}a_{T}+M_{T}B_{T}\overset{\rightsquigarrow}{P}(\Lambda)\\
 & =\vartheta_{T}+\Theta_{T}\overset{\rightsquigarrow}{P}(\Lambda),
\end{align*}
with 
\begin{align}
\vartheta_{T} & =M_{T}a_{T},\label{eq:aux-17}\\
\Theta_{T} & =M_{T}B_{T}\label{eq:aux-18}
\end{align}

2) Now 
\begin{align*}
a_{T} & =\phi_{T}r_{T}+a_{T-1},\\
B_{T} & =\phi_{T}s^{\top}_{T}+B_{T-1},
\end{align*}
with 
\begin{align*}
r_{T} & =\xi^{\top}_{T}\Delta\xi_{T},\\
s_{T} & =\gamma x_{T+1}\otimes x_{T+1}.
\end{align*}
Also, using the matrix inversion lemma, we obtain 
\begin{align}
M_{T} & =M_{T-1}-\frac{M_{T-1}\phi_{T}\phi^{\top}_{T}M_{T-1}}{1-\phi^{\top}_{T}M_{T-1}\phi_{T}}\label{eq:aux-19}\\
 & =M_{T-1}-\eta_{T}\phi^{\top}_{T}M_{T-1}\nonumber 
\end{align}
Hence, from~(\ref{eq:aux-17}) 
\begin{align}
\vartheta_{T} & =M_{T}\phi_{T}r_{T}+M_{T}a_{T-1}\nonumber \\
 & =M_{T}\phi_{T}r_{T}+M_{T-1}a_{T-1}-\eta_{T}\phi^{\top}_{T}M_{T-1}a_{T-1}\nonumber \\
 & =M_{T}\phi_{T}r_{T}+\left(I-\eta_{T}\phi^{\top}_{T}\right)\vartheta_{T-1},\label{eq:aux-20}
\end{align}
and from~(\ref{eq:aux-18}) 
\begin{align}
\Theta_{T} & =M_{T}\phi_{T}s^{\top}_{T}+M_{T}B_{T-1}\nonumber \\
 & =M_{T}\phi_{T}s^{\top}_{T}+M_{T-1}B_{T-1}-\eta_{T}\phi^{\top}_{T}M_{T-1}B_{T-1}\nonumber \\
 & =M_{T}\phi_{T}s^{\top}_{T}+\left(I-\eta_{T}\phi^{\top}_{T}\right)\Theta_{T-1}.\label{eq:aux-21}
\end{align}

3) Using~(\ref{eq:aux-19}) we obtain 
\begin{align*}
M_{T}\phi_{T} & =M_{T-1}\phi_{T}-\frac{M_{T-1}\phi_{T}\phi^{\top}_{T}M_{T-1}\phi_{T}}{1-\phi^{\top}_{T}M_{T-1}\phi_{T}}\\
 & =\frac{M_{T-1}\phi_{T}}{1-\phi^{\top}_{T}M_{T-1}\phi_{T}}\\
 & =\eta_{T}.
\end{align*}
The result then follows by putting the above into~(\ref{eq:aux-20})
and~(\ref{eq:aux-21}). 
\end{IEEEproof}
We now state the complexity of the algorithm presented in Proposition~\ref{prop:RLS-qlearning}.
Letting $D=(N+M)(N+M+1)/2+1$, the complexity per sample time is: 
\begin{itemize}
\item $D-1$ to compute $\phi_{t}$, 
\item $N^{2}+N+M^{2}+M$ to compute $\xi^{\top}_{t}\Delta\xi_{t}$, 
\item $N(N+1)/2$ to compute $x_{T+1}\tilde{\otimes}x_{T+1}$, 
\item $D$ to compute $M_{T-1}\phi_{T}$. 
\item $D$ to compute $M_{T}$, 
\item $2D$ to compute $\eta_{T}$, 
\item $D^{2}$ to compute $I-\eta_{T}\phi^{\top}_{T}$, 
\item $D(D+1)$ to compute $\vartheta_{T},$ 
\item $D(D+1)N(N+1)/2$ to compute $\Theta_{T}$ 
\item $NM^{2}+N\times\mathrm{GAUSS}(M)$ to compute $P(\Lambda)$. 
\item $DN(N+1)/2$ to compute $\theta_{T}$. 
\end{itemize}

\section{Simulations}

\label{sec:Simulations}

In this section we present numerical experiments comparing the performance
of both, the classic approach, which we denote by SysId+LQG and the
proposed approach, which we denote by Q-learning. We do the comparison
in terms of accuracy and complexity. In view of~(\ref{eq:error-covariance}),
we measure the former using 
\[
e_{T}=\mathrm{Tr}\left\{ \mathcal{C}_{T}\right\} ,
\]
where the expected value is estimated by averaging $1000$ Monte Carlo
runs.

In Section~\ref{subsec:Test-system} we use a test system for performance
evaluation. In Section~\ref{subsec:Complexity-comparison} we compare
the complexity of both methods for different system dimensions.

\subsection{Test system}

\label{subsec:Test-system}

In this section we compare the performance of the SysId+LQG and Q-learning
methods using the following system taken from~\cite{Hagen1998} 
\begin{align*}
A & =\left[\begin{array}{cc}
-0.6 & -0.4\\
1 & 0
\end{array}\right], & B & =\left[\begin{array}{c}
0\\
1
\end{array}\right],\\
\Sigma & =\left[\begin{array}{cc}
0.01 & 0\\
0 & 0.01
\end{array}\right], & Q & =\left[\begin{array}{cc}
1 & 0\\
0 & 1
\end{array}\right],\\
R & =\left[1\right], & \gamma & =0.99.
\end{align*}

\begin{figure}
\centering \includegraphics[width=0.8\columnwidth]{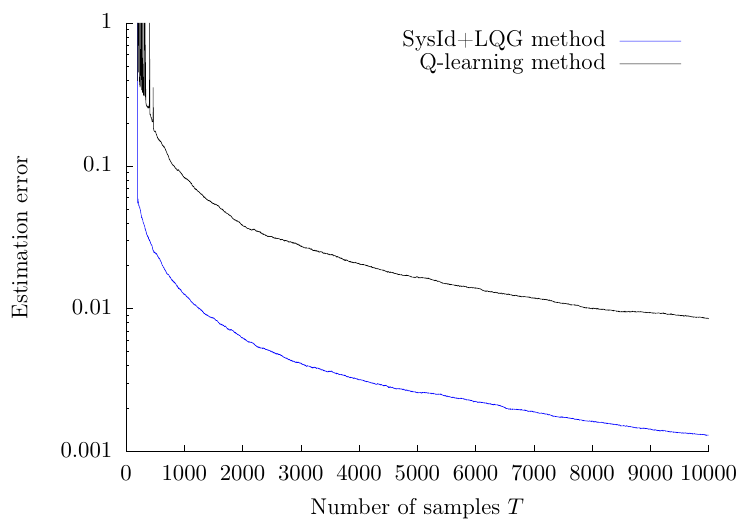}
\caption{Accuracy comparison using a random input.}
\label{fig:comparison-rnd-input} 
\end{figure}

In the first experiment we choose the input by randomly drawing it
from a normal distribution. More precisely, we take 
\begin{equation}
u_{t}\sim\mathcal{N}\left(0,LCL^{\top}\right).\label{eq:input-rnd}
\end{equation}
where $C$ is the steady state covariance of the state $x_{t}$, when
the system runs under the nominal SF-LQG controller~(\ref{eq:control-policy}),
i.e., 
\[
C=(A-BL)C(A-BL)^{\top}+\Sigma.
\]
The result is shown in Figure~\ref{fig:comparison-rnd-input}. As
expected, the SysId+LQG method is more accurate than Q-learning since,
as shown in Theorem~\ref{thm:ML-conv}, it is asymptotically efficient.

\begin{figure}
\centering \includegraphics[width=0.8\columnwidth]{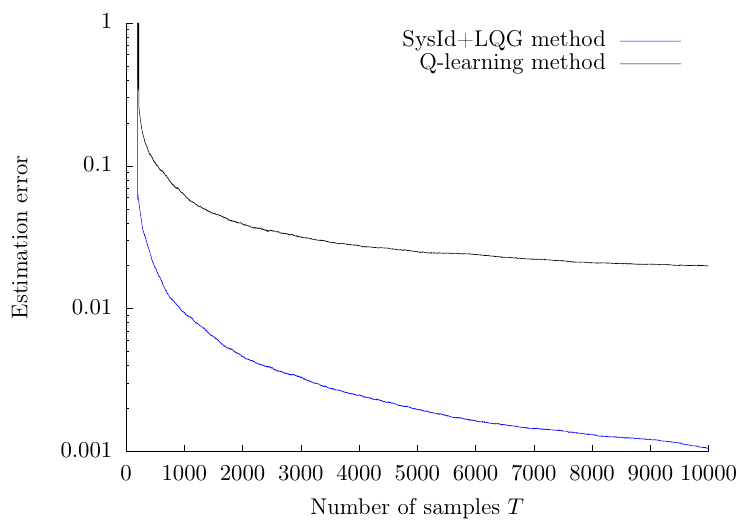}\caption{Accuracy comparison using the LQG input.}
\label{fig:comparison-lqg-input} 
\end{figure}

The reason for choosing a random input policy in the above experiment
is to guarantee that assumptions~(\ref{eq:assu1}) and~(\ref{eq:assu2}),
of Theorems~\ref{thm:ML-conv} and~\ref{thm:qlearning-conv}, respectively,
are satisfied. In the second experiment we drop this policy and choose
the input using~(\ref{eq:input-rnd}) up to $T=200$, and then switch
to the currently estimated SF-LQG controller policy 
\[
u_{t}=-\hat{L}_{t}x_{t}.
\]
The result in shown in Figure~\ref{fig:comparison-lqg-input}. We
see that the above policy significantly slows down the performance
of the Q-learning method, while does not sensibly affect that of the
SysId+LQG method.

Concerning the complexity, the SysId+LQG method requires $130.3$
multiplications per sample time while the Q-learning method requires
$200.7$. Hence, we conclude that the SysId+LQG approach clearly outperforms
the Q-learning one in both, accuracy and complexity.

\subsection{Complexity comparison}

\label{subsec:Complexity-comparison}

\begin{figure}[t]
\centering \includegraphics[width=0.8\columnwidth]{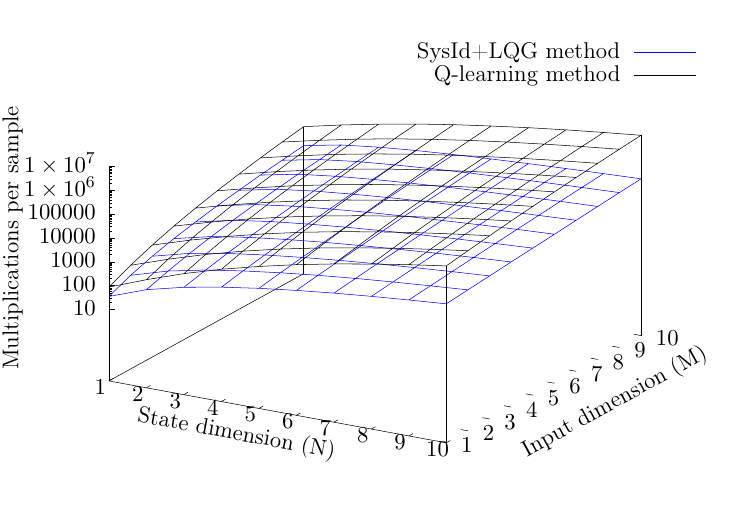}
\caption{Complexity comparison, \emph{logarithmic} scale.}
\label{fig:cc_comparison-log} 
\end{figure}

The complexity of both methods depends on the state dimension $N$
and the input dimension $M$. In Figure~\ref{fig:cc_comparison-log}
we plot the complexities of both methods in logarithmic scale. We
see that the SysId+LQG method is numerically more efficient in the
whole $(N,M)$-plane. Moreover, the difference in complexity between
both methods decreases as $N$ and $M$ decrease. But even for $N=M=1$,
the SysId+LQG requires 34.67 multiplications per sample time, which
is clearly smaller than the complexity of Q-learning which requires
86.33 multiplications per sample time.

\section{Conclusion}

\label{sec:Conclusion}

Our work centers in the problem of designing a SF-LQG controller for
a linear Gaussian system with unknown parameters. We have conducted
a rigorous comparison between the classic approach consisting in using
system identification in combination with a SF-LQG design, with a
recently proposed approach using the Q-learning paradigm. The comparison
is done in terms of both, the accuracy and the complexity of the resulting
controller. In order to build a theoretical ground for this comparison,
we have shown that the classic approach asymptotically achieves the
theoretical lower accuracy bound given by the CRLB, giving virtually
no room for improvement in terms of accuracy. We have also proposed
a novel Q-learning-based method which we show asymptotically achieves
the optimum controller design. We have complemented this method with
a numerical efficient algorithm aiming at making it competitive in
terms of complexity. Nevertheless, our complexity analysis shows that
the classic approach is still computationally more efficient than
this Q-learning-based counterpart. We conclude that the classic approach
remains to be the best option for designing a SF-LQG controller in
the studied scenario.

\appendices{}

\section{Fréchet derivatives}

\label{sec:Frechet derivatives}

The Fréchet derivative generalizes the concept of derivative of functions
between Euclidean spaces to functions between normed vector spaces~ 
\begin{defn}
Let $X$ and $Z$ be normed vector spaces and $U\subseteq X$ be open.
A function $f:U\rightarrow Z$ is called Fréchet differentiable at
$x\in U$ if there exists a bounded linear map $A\in\mathcal{B}(X,Z)$
from $X$ to $Z$ such that 
\[
\lim_{\left\Vert h\right\Vert \rightarrow0}\frac{\left\Vert f(x+h)-f(x)-A(h)\right\Vert }{\left\Vert h\right\Vert }=0.
\]
In this case we say that $A$ is the Fréchet derivative of $f$ at
$x$, and denote it by $\mathfrak{D}f(x)=A$. We also say that $f$
is Fréchet differentiable if it is so at all $x\in U$. We use $\mathfrak{D}f:U\rightarrow\mathcal{B}\left(X,Z\right):x\mapsto\mathfrak{D}f(x)$
to denote the Fréchet derivative of $f$.

If $f:X\times Y\rightarrow Z$, we define the partial Fréchet derivative
$\mathfrak{D}_{x}f(x,y)$ of $f$ with respect to $x$ at $(x,y)$,
as the Fréchet derivative of the map $x\mapsto f(x,y)$ at $x$. 
\end{defn}
As the derivative of functions between Euclidean spaces, Fréchet derivatives
enjoy the following properties. 
\begin{prop}
\label{prop:FD-properties} TFHT 
\begin{enumerate}
\item \label{enu:Linearity}Linearity 
\[
\mathscr{D}(\alpha f+\beta g)(x)=\alpha\mathscr{D}f(x)+\beta\mathscr{D}g(x).
\]
\item \label{enu:Chain-rule}Chain rule 
\[
\mathscr{D}(g\circ f)(x)=\mathscr{D}g\left(f(x)\right)\circ\mathscr{D}f(x).
\]
\end{enumerate}
\end{prop}
We start by providing some Fréchet derivatives which are instrumental
to our goal. 
\begin{lem}
\label{lem:FD-useful} TFHT 
\begin{enumerate}
\item If $f(X)=X^{-1}$, then $\mathfrak{D}f(X)(\mathfrak{M})=-X^{-1}\mathfrak{M}X^{-1}$. 
\item If $f(X)=\log\left|X\right|$, then $\mathfrak{D}f(X)(\mathfrak{M})=\mathrm{Tr}\left\{ X^{-1}\mathfrak{M}\right\} $. 
\item \label{enu:cov}If $\log\left|X\right|+\mathrm{Tr}\left\{ X^{-1}A\right\} $,
then $\mathfrak{D}f(X)(\mathfrak{M})=\mathrm{Tr}\left\{ X^{-1}\mathfrak{M}X^{-1}\left(X-A\right)\right\} $. 
\item \label{enu:trace}If $f(X)=\mathrm{Tr}\{AX\}$, then $\mathfrak{D}f(X)(\mathfrak{M})=\mathrm{Tr}\left\{ A\mathfrak{M}\right\} $. 
\item \label{enu:outer-prod}If $f(x)=xx^{\top}$, then $\mathfrak{D}f(x)(\mathfrak{m})=\mathfrak{m}x^{\top}+x\mathfrak{m}^{\top}$. 
\item \label{enu:affine}If $f(X)=a+Xb$, then $\mathfrak{D}f(X)(\mathfrak{M})=\mathfrak{M}b$. 
\end{enumerate}
\end{lem}
\begin{IEEEproof}
We show each claim separately: 
\begin{enumerate}
\item For a scalar parameter $\alpha$ we have 
\begin{align*}
\frac{\partial X(\alpha)^{-1}}{\partial\alpha} & =-X(\alpha)^{-1}\frac{\partial X(\alpha)}{\partial\alpha}X(\alpha)^{-1},
\end{align*}
Then 
\begin{align*}
\mathfrak{D}f(X)(\mathfrak{M}) & =\mathfrak{D}f\left(\sum_{i,j}[\mathfrak{M}]_{i,j}E_{i,j}\right)\\
 & =\sum_{i,j}[\mathfrak{M}]_{i,j}\mathfrak{D}f\left(E_{i,j}\right)\\
 & =\sum_{i,j}[\mathfrak{M}]_{i,j}\mathfrak{D}_{x_{i,j}}f(X)\left(1\right)\\
 & =\sum_{i,j}[\mathfrak{M}]_{i,j}\frac{\partial X^{-1}}{\partial X_{i,j}}\\
 & =\sum_{i,j}[\mathfrak{M}]_{i,j}\left[-X^{-1}\frac{\partial X}{\partial X_{i,j}}X^{-1}\right]\\
 & =-X^{-1}\left(\sum_{i,j}[\mathfrak{M}]_{i,j}E_{i,j}\right)X^{-1}\\
 & =-X^{-1}\mathfrak{M}X^{-1}.
\end{align*}
\item For a scalar parameter $\alpha$ we have 
\[
\frac{\partial\log\left|X(\alpha)\right|}{\partial\alpha}=\mathrm{Tr}\left\{ X(\alpha)^{-1}\frac{\partial X(\alpha)}{\partial\alpha}\right\} .
\]
Then 
\begin{align*}
\mathfrak{D}f(X)(\mathfrak{M}) & =\mathfrak{D}f(X)\left(\sum_{i,j}[\mathfrak{M}]_{i,j}E_{i,j}\right)\\
 & =\sum_{i,j}[\mathfrak{M}]_{i,j}\mathfrak{D}f(X)\left(E_{i,j}\right)\\
 & =\sum_{i,j}[\mathfrak{M}]_{i,j}\mathfrak{D}_{x_{i,j}}f(x)\left(1\right)\\
 & =\sum_{i,j}[\mathfrak{M}]_{i,j}\frac{\partial\log\left|X\right|}{\partial X_{i,j}}\\
 & =\sum_{i,j}[\mathfrak{M}]_{i,j}\mathrm{Tr}\left\{ X^{-1}\frac{\partial X}{\partial X_{i,j}}\right\} \\
 & =\sum_{i,j}[\mathfrak{M}]_{i,j}\mathrm{Tr}\left\{ X^{-1}E_{i,j}\right\} \\
 & =\mathrm{Tr}\left\{ X^{-1}\sum_{i,j}[\mathfrak{M}]_{i,j}E_{i,j}\right\} \\
 & =\mathrm{Tr}\left\{ X^{-1}\mathfrak{M}\right\} .
\end{align*}
\item We have 
\begin{align}
 & \mathfrak{D}f(X)(\mathfrak{M})\nonumber \\
= & \mathfrak{D}\left\{ \log\left|X\right|\right\} (\mathfrak{M})+\mathrm{Tr}\left\{ \mathfrak{D}\left\{ X^{-1}\right\} (\mathfrak{M})A\right\} \nonumber \\
= & \mathrm{Tr}\left\{ X^{-1}\mathfrak{M}\right\} -\mathrm{Tr}\left\{ X^{-1}\mathfrak{M}X^{-1}A\right\} \nonumber \\
= & \mathrm{Tr}\left\{ X^{-1}\mathfrak{M}X^{-1}\left(X-A\right)\right\} .\label{eq:Q-fder}
\end{align}
\item We have 
\begin{align*}
\mathfrak{D}f(X)(\mathfrak{M}) & =\mathfrak{D}f(X)\left(\sum_{i,j}[\mathfrak{M}]_{i,j}E_{i,j}\right)\\
 & =\sum_{i,j}[\mathfrak{M}]_{i,j}\mathfrak{D}f(X)\left(E_{i,j}\right)\\
 & =\sum_{i,j}[\mathfrak{M}]_{i,j}\mathfrak{D}_{x_{i,j}}\left\{ f(X)\right\} \left(1\right)\\
 & =\sum_{i,j}[\mathfrak{M}]_{i,j}\frac{\partial\mathrm{Tr}(AX)}{\partial X_{i,j}}\\
 & =\sum_{i,j}[\mathfrak{M}]_{i,j}\frac{\partial\sum_{k,l}A_{k,l}X_{l,k}}{\partial X_{i,j}}\\
 & =\sum_{i,j}[\mathfrak{M}]_{i,j}A_{j,i}\\
 & =\mathrm{Tr}\left\{ A\mathfrak{M}\right\} .
\end{align*}
\item We have 
\begin{align*}
\mathfrak{D}f(x)(\mathfrak{m}) & =\mathfrak{D}f(x)\left(\sum_{i}[\mathfrak{m}]_{i}e_{i}\right)\\
 & =\sum_{i}[\mathfrak{m}]_{i}\mathfrak{D}f(x)\left(e_{i}\right)\\
 & =\sum_{i}[\mathfrak{m}]_{i}\mathfrak{D}_{x_{i}}\left\{ f(x)\right\} \left(1\right)\\
 & =\sum_{i}[\mathfrak{m}]_{i}\frac{\partial xx^{\top}}{\partial x_{i}}\\
 & =\sum_{i}[\mathfrak{m}]_{i}\left(e_{i}x^{\top}+xe^{\top}_{i}\right)\\
 & =\mathfrak{m}x^{\top}+x\mathfrak{m}^{\top}.
\end{align*}
\item We have 
\begin{align*}
\mathfrak{D}f(X)(\mathfrak{M}) & =\mathfrak{D}f(X)\left(\sum_{i,j}[\mathfrak{M}]_{i,j}E_{i,j}\right)\\
 & =\sum_{i,j}[\mathfrak{M}]_{i,j}\mathfrak{D}f(X)\left(E_{i,j}\right)\\
 & =\sum_{i,j}[\mathfrak{M}]_{i,j}\mathfrak{D}_{x_{i,j}}\left\{ f(X)\right\} \left(1\right)\\
 & =\sum_{i,j}[\mathfrak{M}]_{i,j}\frac{\partial\left(a+Xb\right)}{\partial X_{i,j}}\\
 & =\sum_{i,j}[\mathfrak{M}]_{i,j}b_{j}e_{i}\\
 & =\mathfrak{M}b.
\end{align*}
\end{enumerate}
\end{IEEEproof}

\section{Generalized chi-squared distribution }

\label{sec:Generalized-chi-squared-distribu} 
\begin{defn}
\label{def:generalized-chi-squared}Let $x\sim\mathcal{N}\left(0,\mathbf{I}\right)$,
$0<A\in\mathbb{R}^{N\times N}$, $b\in\mathbb{R}^{N}$ and $c\in\mathbb{R}$.
Let 
\[
y=x^{\top}Ax+b^{\top}x+c.
\]
We then say that $y$ has generalized chi-squared distribution with
parameters $A$, $b$ and $c$, and denote it by 
\[
y\sim\bar{\chi}\left(A,b,c\right).
\]
\end{defn}
The probability density and cumulative distribution functions of the
generalized chi-squared distribution do not have closed-form expressions.
The mean and variance, however, do have. 
\begin{prop}
\label{prop:GCS-mean-var}In the notation of Definition~\ref{def:generalized-chi-squared},
\begin{align*}
\mathcal{E}\left\{ y\right\}  & =\mathrm{Tr}\left(A\right)+c\\
\mathcal{V}\left\{ y\right\}  & =2\mathrm{Tr}\left(A^{2}\right)+b^{\top}b.
\end{align*}
\end{prop}
\begin{IEEEproof}
We have 
\begin{align*}
\mathcal{E}\left\{ y\right\}  & =\mathcal{E}\left\{ x^{\top}Ax\right\} +c\\
 & =\mathrm{Tr}\left(A\mathcal{E}\left\{ xx^{\top}\right\} \right)+c\\
 & =\mathrm{Tr}\left(A\right)+c.
\end{align*}
Also, let $A=U\Lambda U^{\top}$ be the eigen-decomposition of $A$
and $z=U^{\top}x$. It then follows that $z\sim\mathcal{N}\left(0,\mathbf{I}\right)$
and 
\begin{align*}
y & =z^{\top}\Lambda z+\beta^{\top}z+c\\
 & =c+\sum^{N}_{n=1}\chi_{n},
\end{align*}
with 
\begin{align*}
\beta & =U^{\top}b,\\
\chi_{n} & =\Lambda_{n,n}z^{2}_{n}+\beta_{n}z_{n}.
\end{align*}
Then 
\begin{align*}
\mathcal{V}\left\{ y\right\}  & =\sum^{N}_{n=1}\mathcal{V}\left\{ \chi_{n}\right\} \\
 & =\sum^{N}_{n=1}\left(\mathcal{E}\left\{ \chi^{2}_{n}\right\} -\mathcal{E}^{2}\left\{ \chi_{n}\right\} \right)
\end{align*}
Now 
\begin{align*}
\mathcal{E}\left\{ \chi_{n}\right\}  & =\Lambda_{n,n},\\
\mathcal{E}\left\{ \chi^{2}_{n}\right\}  & =\mathcal{E}\left\{ \Lambda^{2}_{n,n}z^{4}_{n}+2\Lambda_{n,n}\beta_{n}z^{3}_{n}+\beta^{2}_{n}z^{2}_{n}\right\} \\
 & =3\Lambda^{2}_{n,n}+\beta^{2}_{n}.
\end{align*}
Hence 
\begin{align*}
\mathcal{V}\left\{ y\right\}  & =\sum^{N}_{n=1}\left(2\Lambda^{2}_{n,n}+\beta^{2}_{n}\right)\\
 & =2\sum^{N}_{n=1}\lambda^{2}_{n}(A)+b^{\top}UU^{\top}b\\
 & =2\mathrm{Tr}\left(A^{2}\right)+b^{\top}b.
\end{align*}
\end{IEEEproof}
 \bibliographystyle{IEEEtran}
\bibliography{bibliophaphy}

\end{document}